\documentclass{pas}

\usepackage{multirow,newtxtext,newtxmath,mathtools,amsmath,gensymb,makecell,subcaption,float,threeparttable}
\usepackage{stfloats}
\usepackage{url}      

\begin{document}

\lefttitle{Publications of the Astronomical Society of Australia}
\righttitle{M. Mousavi-Sadr et al.}

\jnlPage{1}{4}
\jnlDoiYr{2021}
\doival{10.1017/pasa.xxxx.xx}

\articletitt{Research Paper}

\title{Exoplanetary radio emission predictions and detectability in the SKA era}

\author{\sn{M.} \gn{Mousavi-Sadr}$^{1,2}$, \sn{F.~S.} \gn{Tabatabaei$^{1}$}, \sn{A.} \gn{Wolszczan$^{3,4}$}, and \sn{G.} \gn{Gozaliasl$^{5,6}$}}

\affil{$^1$School of Astronomy, Institute for Research in Fundamental Sciences (IPM), P. O. Box 19395-5531, Tehran, Iran, $^{2}$Iranian National Observatory (INO), Institute for Research in Fundamental Sciences (IPM), P. O. Box 19568-36613, Tehran, Iran, $^3$Department of Astronomy and Astrophysics, Pennsylvania State University, 525 Davey Laboratory, University Park, PA 16802, USA, $^{4}$Center for Exoplanets and Habitable Worlds, Pennsylvania State University, 525 Davey Laboratory, University Park, PA 16802, USA, $^{5}$Department of Computer Science, Aalto University, P. O. Box 15400, Espoo, FI-00076, Finland, $^{6}$Department of Physics, University of Helsinki, P. O. Box 64, FI-00014, Helsinki, Finland}

\corresp{M. Mousavi-Sadr, Email: mahdiyar@ipm.ir}

\citeauth{M. Mousavi-Sadr, Exoplanetary radio emission predictions and detectability in the SKA era. {\it Publications of the Astronomical Society of Australia} {\bf 00}, 1--12. https://doi.org/10.1017/pasa.xxxx.xx}

\history{(Received xx xx xxxx; revised xx xx xxxx; accepted xx xx xxxx)}

\begin{abstract}
Radio observations provide a window into a planet’s interior and play a crucial role in studying its atmosphere and surface, key factors to find potential habitability. The discovery of thousands of exoplanets, together with advances in radio astronomy through the Square Kilometre Array (SKA), motivates the search for planetary-scale radio emissions. Here, we employ the radiometric Bode’s law (RBL) to analyze a data set of 1027 confirmed exoplanets orbiting F-M main-sequence dwarf stars to estimate their potential auroral radio emission. Permutation Importance (PI) and SHapley Additive exPlanations (SHAP) analyses indicate that planetary radius, orbital semi-major axis, stellar age, and distance from Earth are sufficient to dependably forecast a planet's radio flux density. Considering observational constraints, we find that 15 exoplanets could generate signals detectable by the SKA, nine of which remain observable in the intermediate AA* deployment. Among these, WASP-14 Ab stands out with a predicted flux of 40.064~mJy at 57.5~MHz, making it an excellent SKA-Low target. Meanwhile, HD 141937 b, with a flux of 0.084~mJy peaking at 733.1~MHz, is the most promising candidate for SKA-Mid. These results show that the SKA can detect gas giants, such as WASP-14 Ab (SNR$>$245) and HD 141937 b (SNR$>$19), within feasible integration times. Additionally, we identify one candidate (TOI-4515 b) that is likely affected by radio quenching, highlighting the importance of considering this effect in target selection for observation campaigns.
\end{abstract}


\maketitle

\section{Introduction}
In the solar system, Earth and the gas giants generate non-thermal radio emissions through cyclotron maser instability (CMI). This process occurs when a beam of electrons from the solar wind spirals around a planet's magnetic field lines, producing low-frequency, coherent, and circularly polarised radio signals \citep{1999JGR...10414025F,2018haex.bookE..22Z,2024arXiv240412348L,2025ARA&A..63..299V}. Emitted radio frequencies correspond to the local cyclotron frequency, which depends on the planet's magnetic field strength and ranges from a few kHz far from the planet up to the maximum surface cyclotron frequency ($f_{c}$). For most solar system planets, the surface magnetic field strength is less than 1 G, resulting in emitted radio frequencies typically below $\sim$2 MHz. These low-frequency signals are reflected by the Earth's ionosphere, making them undetectable from the ground. Jupiter, however, has the strongest magnetic field among the solar system planets, with a surface field strength of up to 14 G, and emits radio bursts at frequencies up to 40 MHz. Ground-based radio telescopes can easily detect these bursts because they fall within the observable radio frequency range ($>$10 MHz) \citep{1999JGR...10414025F,2015aska.confE.120Z}.

Like the planets in the solar system, exoplanets with sufficiently strong magnetic fields are also expected to produce auroral radio emission through the CMI mechanism. Detecting the radio emission from exoplanets can provide a wealth of information that might be impossible to obtain through other methods.

Planetary masses and radii serve as key parameters for estimating a planet's interior structure and composition. Nevertheless, determining the internal structure of a planet is highly challenging, as different compositions can result in the same mass and radius \citep{2010ApJ...716.1208R,2014ApJ...792....1L,2017A&A...597A..37D,2018ApJ...866...49L,2020A&A...640A.135O}. Radio detection is likely the only method that can clearly reveal an exoplanet's magnetic field, which, in turn, can offer insights into its internal composition while also helping to resolve this intrinsic degeneracy \citep{2015ASSL..411..213G,2018haex.bookE.159G}.

Moreover, the presence of a planetary magnetosphere is a crucial factor in the potential development of life. It protects a potentially habitable planet from severe cellular damage and genetic material disruption in any life on its surface by deflecting incoming cosmic rays. A planet's magnetic field can also shield its atmosphere and prevent its erosion by the stellar wind or coronal mass ejections (CMEs) \citep{2004ApJ...612..511L,2015aska.confE.120Z}.

Neptune's rotation period was first estimated based on its differentially rotating cloud tops but was later refined through observations of its radio emission. Since a planet's magnetic field is linked to its interior, it offers a more precise measure of the planetary rotation rate than atmospheric features like cloud motions. Likewise, periodic radio emission of exoplanets can be used to determine their rotation periods \citep{1993GeoRL..20.2711L,2004ApJ...612..511L}. Ultimately, radio detection may also emerge as a new method for discovering exoplanets, complementing transit and radial velocity measurements, as it is particularly well-suited for detecting planets around active, magnetic, or variable stars \citep{2015aska.confE.120Z,2024arXiv240412348L}.

Numerous studies have been conducted to estimate exoplanetary radio emissions. The first predictive efforts were carried out by \cite{1997pre4.conf..101Z} and \cite{1999JGR...10414025F}. At that time, only a few dozen exoplanets had been discovered, so their studies focused on a limited sample. As the number of exoplanet discoveries grew, the first catalogue of estimated exoplanetary radio emissions was presented by \cite{2004ApJ...612..511L}, applying the radiometric Bode’s law (RBL) to 118 exoplanets across 102 systems. Their estimations suggested that most of the exoplanets in the catalogue would emit at frequencies between 10 and 1000 MHz, with flux densities potentially reaching up to 1 mJy. They also conducted a systematic search for radio emission using low-frequency images from the Very Large Array (VLA). Although no significant signals were detected, an upper limit of 300 mJy was established, which was consistent with model predictions but did not place strong constraints on them.

Moreover, in a detailed analysis, \cite{2007A&A...475..359G} compared the radio emission predictions of various theoretical models for all 220 known exoplanets at the time. They found that different models yield widely varying results for planetary radio emission. The authors concluded that while the detection of exoplanetary radio emission is feasible, the number of promising targets was relatively limited at that time \citep{2007yCat..34750359G}.

In an observational study, \cite{2018A&A...612A..52O} investigated the potential for detecting exoplanetary radio emission by targeting evolved stars, where high stellar mass-loss rates could facilitate the generation of detectable planetary signals. Using the RBL model, they selected three evolved star systems including $\beta$~Gem, $\iota$~Dra, and $\beta$~UMi as promising candidates and conducted radio observations at 150 MHz with the LOw Frequency ARray (LOFAR). Although no radio emission was detected, 3$\sigma$ upper limits were placed on the flux densities, demonstrating LOFAR's capabilities and highlighting the promising role of the upcoming SKA Observatory in searching for exoplanetary radio emissions at meter wavelengths.

Recently, the feasibility of detecting radio emissions from 83 exoplanets using the SKA was investigated by \cite{2024FrASS..1112323B}. Employing the RBL model to estimate the expected flux density, they identified four exoplanets, namely Qatar-4 b, TOI-1278 b, CoRoT-10 b, and HAT-P-20 b, as promising candidates that may produce detectable radio signals with the SKA telescope.

Extensive radio observations targeting exoplanetary emission have been conducted, with non-detections reported from VLA for the hot Jupiter $\tau$ Boo at 74 MHz (100 mJy sensitivity) and HD 80606 at 330 MHz and 1465 MHz (1.7 mJy and 0.048 mJy, respectively) \citep{2003ASPC..294..151F, 2004IAUS..213...73F, 2007ApJ...668.1182L, 2010AJ....140.1929L}. Likewise, the Giant Metrewave Radio Telescope (GMRT) at 150 MHz detected no signal from $\tau$ Boo, with a sensitivity limit of $\sim$1 mJy \citep{2013ApJ...762...34H}. The Ukrainian T-shaped Radio telescope, second modification (UTR-2), also failed to detect signals from a dozen targets, each observed for several hours across the frequency range of $\sim$15 to 30 MHz \citep{2004P&SS...52.1479R}.

Despite these challenges, some encouraging tentative detections have emerged, such as a circularly polarised bursty emission from $\tau$ Boo observed with LOFAR \citep{2021A&A...645A..59T,2024A&A...688A..66T}, a highly polarised periodic emission from Proxima Centauri system \citep{2021A&A...645A..77P}, coherent radio bursts from YZ Ceti b \citep{2023NatAs...7..569P}, and highly circularly polarised bursts from $\beta$ Pic b \citep{2026arXiv260916720O}. Furthermore, many other studies and observational efforts have been undertaken to estimate exoplanetary radio emissions, assess their detectability with various radio telescopes, and attempt direct detection of these signals. These include, but are not limited to, \cite{2011RaSc...46.0F09G}, \cite{2016MNRAS.461.2353N}, \cite{2018MNRAS.478.1763L}, \cite{2018ApJ...854...72T}, \cite{2019RAA....19...23Z}, \cite{2022BAAS...54e.303A}, \cite{2023pre9.conf03092M}, \cite{2023arXiv231005363T}, and \cite{2024AJ....168..127O}.

Given the difficulties in achieving unambiguous detections of exoplanetary radio emission, as evidenced by prior campaigns, advanced instruments like the SKA and next-generation VLA will be crucial for securing definitive results. The SKA Observatory will consist of two telescopes: SKA-Low and SKA-Mid. SKA-Low, located in Western Australia, will consist of 131,072 dipole antennas distributed across 512 stations and will operate in the 50-350 MHz frequency range. SKA-Mid, located in South Africa’s Karoo region, will consist of 133 15-meter SKA dishes and 64 13.5-meter MeerKAT dishes, covering a wide frequency range from 350 MHz to 50 GHz. Nevertheless, the frequency range between 1.76 and 4.6 GHz as well as above 15.4 GHz are not included in the current deployment plan \citep{2009IEEEP..97.1482D,2016SPIE.9906E..28W}. The construction and commissioning of the SKA Observatory are organized into incremental stages called Array Assemblies (AAs). It is currently progressing toward an intermediate milestone called AA*, which includes 307 Low stations and 144 Mid dishes, a partial deployment compared to the full design baseline known as AA4 \citep{2018AAS...23115207B,2019arXiv191212699B}.

The SKA is set to play a transformative role in planetary science. With its high sensitivity, broad frequency coverage, and large dynamic range, it emerges as one of the most promising instruments for detecting the first-ever radio signal from an exoplanet \citep{2015aska.confE.174B}. In this study, we estimate the expected radio emission from a large sample of confirmed exoplanets using semi-empirical models and introduce a novel approach for identifying the physical parameters that most strongly influence their estimated radio-emission properties. Furthermore, considering the SKA's sky coverage, frequency range, and sensitivity at the AA4 and AA* stages, we identify a subset of planets that could be detectable with this telescope.

This paper is organised as follows: section~\ref{data} describes the data sample. Section~\ref{meth_resu} describes the methods used to predict the planetary radio emission and assess the relative importance of the input parameters in the predictions, together with the corresponding results. Section~\ref{ska} provides the list of potentially detectable exoplanets with the SKA. Section~\ref{disc} examines the radio-quenching effect, discusses the assumptions and uncertainties underlying the RBL framework, addresses relevant observational considerations, and compares our predictions with observational campaigns. Finally, section~\ref{conclusion} summarises the main findings and outlines future directions for observations with the SKA.

\begin{figure}
    \centering
    \includegraphics[width=1\columnwidth]{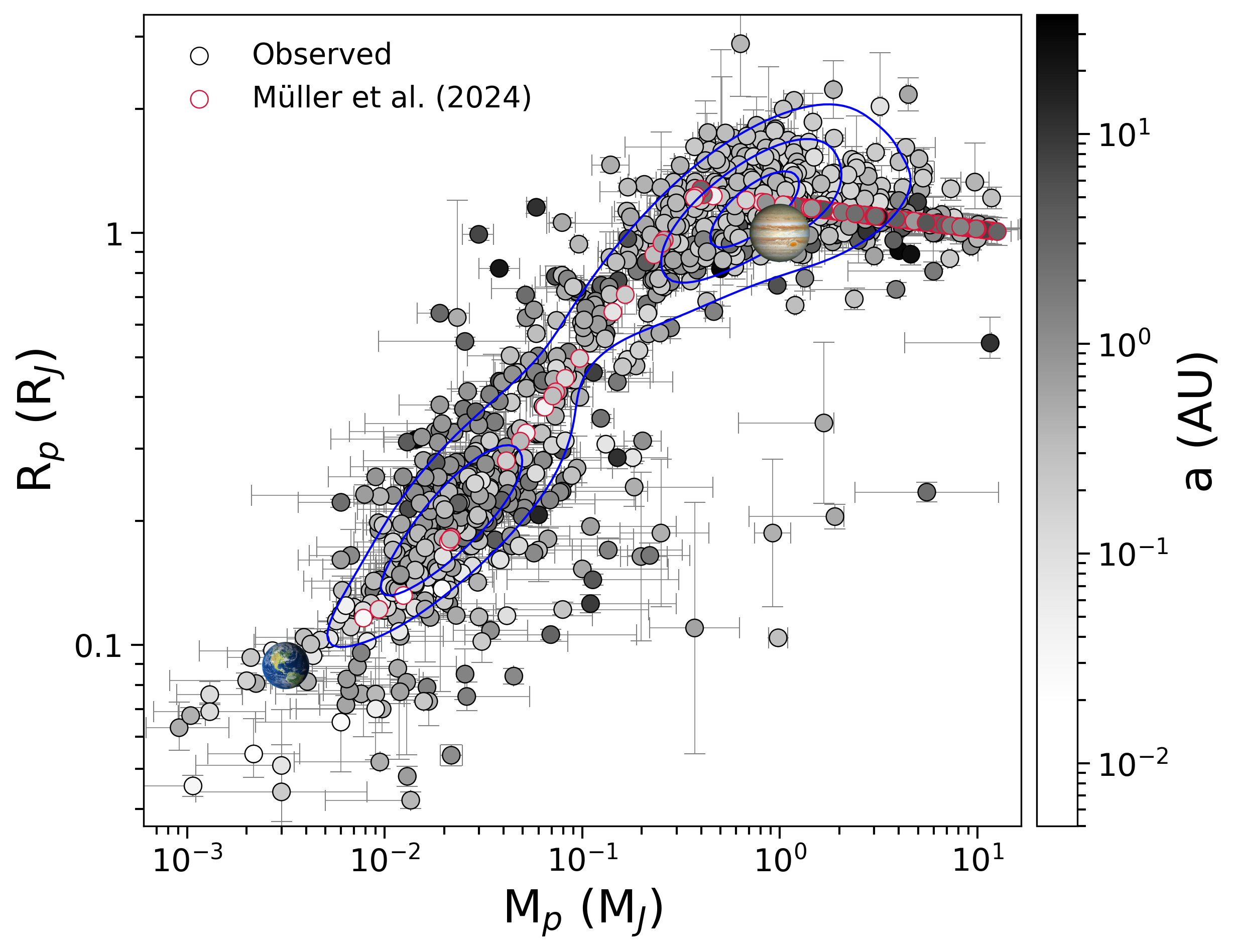}
    \caption{The planetary radius ($R_{p}$) plotted as a function of planet's mass ($M_{p}$) and colour-coded by orbital semi-major axis ($a$), for a sample of 1027 planets. Planets with radii determined using the mass-radius relation from \citet{2024A&A...686A.296M} (see equation~\ref{eq1}) are highlighted with red-bordered circles. These data points follow the mass-radius trend seen among the observed planetary samples. Blue lines represent kernel density estimate contours. The positions of Earth and Jupiter within this parameter space, as two radio-emitting planets in the solar system, are depicted.}
    \label{fig1}
\end{figure}

\section{Data set}\label{data}
We use the Extrasolar Planets Encyclopedia\footnote{\url{http://exoplanet.eu/}} to extract data on exoplanetary systems. To obtain a homogeneous sample, we exclude objects with masses above the canonical deuterium-burning limit of $\sim$13 $M_{J}$ and those with masses below 0.05~$M_{\oplus}$, leaving 6466 confirmed exoplanets as of 8 August 2026. We restrict our analysis to exoplanets orbiting main-sequence dwarf stars of spectral types F, G, K, and M, with stellar ages exceeding 0.7 Gyr, consistent with the stellar-wind models adopted to estimate the radio emission properties (see section~\ref{rbl}), and exclude all other systems. Of the remaining systems, 1027 exoplanets possess parameters, including planetary mass ($M_{p}$), stellar mass ($M_{s}$), radius ($R_{s}$), effective temperature ($T_\text{eff}$) and age ($t$), and the distance to the host star ($D$). Among these, 109 planets have primarily been discovered using radial velocity, direct imaging, and transit timing variation methods, which do not provide measurements of the planet's radius ($R_{p}$). To address this, we apply the following mass-radius relation derived by \cite{2024A&A...686A.296M} to these 109 planets:
\begin{equation}
R_{p}=
\begin{cases}
(1.02\pm0.03)M_{p}^{(0.27\pm0.04)} & M_{p}<(4.37\pm0.72) \\
(0.56\pm0.03)M_{p}^{(0.67\pm0.05)} & (4.37\pm0.72)<M_{p}<(127\pm7) \\
(18.6\pm6.7)M_{p}^{(-0.06\pm0.07)} & M_{p}>(127\pm7),
\end{cases}
\label{eq1}
\end{equation}

where $R_p$ and $M_p$ are in Earth units. Furthermore, 59 planets in the sample do not have a reported semi-major axis ($a$). For these planets, we determine $a$ using Kepler's third law:
\begin{equation}
a=\bigg[\frac{G(M_{s}+M_{p})P_{\text{orb}}^{2}}{4\pi^{2}}\bigg]^{1/3},
\label{eq2}
\end{equation}
where $P_{\text{orb}}$ is the planet's orbital period. Therefore, we construct a data set consisting of 1027 confirmed exoplanets, for which the seven parameters ($M_{p}$, $R_{p}$, $M_{s}$, $R_{s}$, $t$, $a$, and $D$) required for calculating planetary radio properties using the RBL model (see section~\ref{rbl}) are available, either from observations or estimated using equations~\ref{eq1} and~\ref{eq2}.

Figure~\ref{fig1} illustrates the mass-radius distribution of planets in our data set, with colours representing the orbital semi-major axis. Red-bordered circles indicate planets whose radii have been derived using the mass-radius relation from \cite{2024A&A...686A.296M}, following the trend seen among the observed planetary samples. It covers a wide range of planetary masses and radii, with approximately 93\% having an orbital semi-major axis less than 1 AU. This data set, like other observational data sets, may be affected by detection biases. It consists of exoplanets that have mostly been discovered by transit and radial velocity, suffering from the detection limits of both methods. Therefore, we caution that this data set does not provide a complete census of the occurrence of planets.

\section{Radio emission from exoplanets}\label{meth_resu}
In this section, we first employ an analytical approach based on the RBL to estimate the auroral radio flux density and characteristic emission frequency of the exoplanetary systems in our sample. These analytical predictions are then used as the basis for a machine-learning analysis, in which we apply Permutation Importance (PI) and SHapley Additive exPlanations (SHAP), two established techniques for assessing feature importance, to investigate the relative contributions of the physical parameters to the calculated radio emission. This sensitivity analysis allows us to identify the physical properties that most strongly influence the predicted radio flux and characteristic emission frequency and to assess which parameters are most relevant for identifying potentially bright exoplanetary radio sources.

\begin{table*}
\centering
\caption{Radio emission properties of exoplanets estimated using the radiometric Bode's law (RBL). Columns 1 and 2 show the row number and the planet's name. Column 3 presents the system's declination. Columns 4-6 display the planet's radius, mass, and orbital semi-major axis. Columns 7-11 are the host star's radius, mass, effective temperature, age, and distance from Earth. The maximum emission frequency and flux density of the planets are listed in columns 12 and 13, respectively.}
\begin{threeparttable}
\begin{tabular}{llllllllllllll}
\hline
\# & Name & \makecell[l]{$\delta$\\($\degree$)} & \makecell[l]{$R_p$\\($R_J$)} & \makecell[l]{$M_p$\\($M_J$)} & \makecell[l]{a\\(AU)} & \makecell[l]{$R_s$\\($R_\odot$)} & \makecell[l]{$M_s$\\($M_\odot$)} & \makecell[l]{$T_{\text{eff}}$\\(K)} & \makecell[l]{t\\(Gyr)} & \makecell[l]{D\\(pc)} & \makecell[l]{$f_c$\\(MHz)} & \makecell[l]{$\Phi$\\(mJy)}\\
\hline
1 & 14 Her b & 43.8  & 1.0$^{\S}$ & 8.9 & 2.84  & 0.7 & 0.9 & 5311 & 5.1 & 18.1 & 409.8 & 0.012 \\
2 & 14 Her c & 43.8  & 1.0$^{\S}$ & 7.9 & 20.00 & 0.7 & 0.9 & 5311 & 5.1 & 18.1 & 306.5 & 0.001 \\
3 & 23 Lib b  & -25.3 & 1.1$^{\S}$ & 1.8 & 0.81  & 1.3 & 1.1 & 5740 & 9.7 & 22.2 & 48.1  & 0.036 \\
4 & 23 Lib c  & -25.3 & 1.1$^{\S}$ & 1.9 & 5.80  & 1.3 & 1.1 & 5740 & 9.7 & 22.2 & 54.9  & 0.003 \\
5 & 51 Peg b  & 20.8  & 1.2 & 0.6 & 0.05  & 1.2 & 1.0 & 5758 & 4.0 & 15.5 & 0.6   & 51.870 \\
\hline
\end{tabular}
\begin{tablenotes}
\item A machine-readable format of this table in its entirety (1027 exoplanets) is available in the online journal.
\item$^{\S}$The planetary radius has been derived using equation~\ref{eq1}, which may introduce additional uncertainties into the predicted radio emission properties (see section~\ref{uncr}).
\end{tablenotes}
\end{threeparttable}
\label{tab1}
\end{table*}

\subsection{Radiometric Bode's law}\label{rbl}
The Earth and gas giant planets emit radio signals through interactions between their magnetic fields and the incoming solar wind \citep{1999JGR...10414025F}. This auroral emission can be estimated and extended to exoplanetary systems using a semi-empirical relationship known as the RBL \citep{2004ApJ...612..511L,2010ASPC..430..175Z}. This model establishes a relationship between a planet's radio emission power, the strength of its magnetic field, and the stellar wind's incident power. The radio flux incident at Earth ($\Phi$) and the characteristic emission frequency ($f_c$) of an exoplanet are obtained from the following equations:
\begin{equation}
\begin{split}
\Phi = 4.6\,\text{mJy}\,\Big(\frac{\omega}{\omega_{J}}\Big)^{-0.2}\Big(\frac{M_{p}}{M_{J}}\Big)^{-0.33}\Big(\frac{R_{p}}{R_{J}}\Big)^{3}\Big(\frac{\Omega}{1.6\,\text{sr}}\Big)^{-1}\\\Big(\frac{D}{10\,\text{pc}}\Big)^{-2}\Big(\frac{n(a)}{1.34\times10^{4}\,\text{cm}^{-3}}\Big)^{0.8}\Big(\frac{v_{\text{eff}}}{100\,\text{km s}^{-1}}\Big)^{2.8};
\end{split}
\label{eq_flux}
\end{equation}

\begin{equation}
f_{c}=23.5\,\text{MHz}\,\Big(\frac{\omega}{\omega_{J}}\Big)\Big(\frac{M_{p}}{M_{J}}\Big)^{5/3}\Big(\frac{R_{p}}{R_{J}}\Big)^{-3},
\label{eq_freq}
\end{equation}
where $\omega$ is the planet's rotation rate, $M_{p}$ and $R_{p}$ denote the planet's mass and radius, respectively, $\Omega$ is the beaming solid angle of the emission, $D$ is the Earth-star distance, $n(a)$ and $v_\text{eff}$ represent the stellar-wind number density and effective velocity at the planetary orbital distance, respectively. This model assumes a Blackett type scaling law to estimate the planet's magnetic moment via $\mu_{B}\propto\omega M_{p}^{5/3}$ \citep{1947Natur.159..658B,2004ApJ...612..511L}. The derivation of these expressions is presented in appendix~\ref{ap1}.

Planets in close-in orbits around their host stars experience strong tidal forces, resulting in gravitational locking, where their orbital period ($P_{\rm orb}$) is equal to their rotation period \citep[$P_{\omega}$,][]{2007A&A...475..359G}. Thus, we assign $\omega=2\pi/P_{\text{orb}}$ for planets with orbits smaller than 0.1 AU, while for the remaining planets, we determine $\omega$ using the Darwin-Radau relation \citep{1999ssd..book.....M}:
\begin{equation}
\omega=\sqrt{\frac{FGM_{p}}{R^{3}_{\text{eq}}}[2.5(1-1.5C)^{2}+0.4]},
\label{eq_omega}
\end{equation}
where $G$ is the gravitational constant. $R_{\text{eq}}$ is the planet's equatorial radius, which we assume to be equivalent to the planet's radius. Although exoplanets exhibit considerable diversity, we adopt a simplified classification based on their bulk density to estimate the parameter $C$: planets with densities above 4~g/cm$^{3}$ are categorised as rocky ($C=0.4$), while those with densities below this threshold are classified as gas giants \citep[$C=0.25$,][]{2018exha.book.....P}. $F$ represents the planet's oblateness, where for giant planets, we adopt Jupiter's oblateness ($F=0.06487$), and for rocky planets, we use the Earth's oblateness ($F=0.00335$).

In principle, as the stellar-wind properties directly determine the incident energy available for planetary radio emission, the adopted stellar-wind model can significantly affect the predicted radio flux densities. At small orbital distances, the stellar wind is still accelerating and has not yet reached the quasi-asymptotic velocity regime \citep{2007P&SS...55..618G}. Using a constant asymptotic wind velocity could consequently overestimate the expected radio emission from close-in planets, particularly hot Jupiters. To account for the radial evolution of the stellar wind, we therefore adopt Parker's isothermal solution to determine the wind velocity as a function of distance from the host star \citep{1958ApJ...128..664P}.

The stellar wind velocity at $1$~AU is determined from the age-dependent relation
\begin{equation}
v(1\,\mathrm{AU},t) = v_{0}\left(1+\frac{t}{\tau}\right)^{-0.43},
\label{eq_vn1au}
\end{equation}
and the particle density at 1~AU is obtained by
\begin{equation}
n(1\,\mathrm{AU},t) = n_{0}\left(1+\frac{t}{\tau}\right)^{-1.86},
\label{eq_vn1au2}
\end{equation}
where $v_{0}=3971\mathrm{km/s}$, $n_{0}=1.04\times10^{11}\mathrm{m^{-3}}$, and $\tau=2.56\times10^{7}\mathrm{yr}$ \citep{1980asfr.symp.....P,2007P&SS...55..618G}. Note that these relations are applicable only to main-sequence stars with ages $t \geq 0.7$ Gyr \citep{2005ApJ...628L.143W}. Scaling the solar-calibrated mass loss rate according to the host star's surface area gives the stellar ionised mass-loss rate
\begin{equation}
\dot M_{\text{ion}}(t) = 4\pi a_{0}^{2}\,n(1\,\mathrm{AU},t)\,v(1\,\mathrm{AU},t)\,m_{p}\left(\frac{R_{s}}{R_{\odot}}\right)^{2},
\label{eq_mdot}
\end{equation}
with $a_{0}=1$~AU and $m_{p}$ the proton mass. Variations in mass-loss rates among stars may introduce statistical fluctuations in the predicted radio power levels, though these effects are expected to be less significant than those arising from differences in stellar wind velocities \citep{2004ApJ...612..511L}. The stellar-wind velocity $v(a)$ at the planet's orbital distance $a$ then follows from Parker's isothermal wind equation \citep{1958ApJ...128..664P,1999A&A...348..614M},
\begin{equation}
\frac{v(a)^{2}}{v_{c}^{2}}-\ln\left(\frac{v(a)^{2}}{v_{c}^{2}}\right)=4\ln\left(\frac{a}{r_{c}}\right)+4\frac{r_{c}}{a} - 3,
\label{eq_parker}
\end{equation}
with the critical velocity $v_{c}=\sqrt{2k_{B}T_{c}/m_{p}}$ and critical radius $r_{c}=m_{p}GM_{s}/4k_{B}T_{c}$. Here, $k_{B}$ is the Boltzmann constant and $T_{c}$ is the coronal temperature. As the coronal temperature is not directly measured, we adjust $T_{c}$ until the stellar-wind velocity at 1~AU obtained from Parker's solution (equation~\ref{eq_parker}) agrees with the value calculated from equation~\ref{eq_vn1au}. We then use the same $T_{c}$ to evaluate the wind velocity at the planet's semi-major axis.

The particle density $n(a)$ at the planet's orbital distance $a$ is then obtained from mass conservation in the steady, spherically symmetric wind:
\begin{equation}
n(a)=\frac{\dot M_{\text{ion}}}{4\pi a^{2}m_{p}v(a)}.
\label{eq_na}
\end{equation}

For planets orbiting close to their host stars, the planetary Keplerian velocity can be comparable to the stellar-wind velocity. These close-in orbits are expected to be approximately circular due to tidal dissipation, in which case the Keplerian velocity is given by
\begin{equation}
\nu_{k}(a)=\sqrt{\frac{GM_s}{a}}.
\label{eq_kep}
\end{equation}
Finally, the effective stellar-wind velocity relative to the planet's orbital motion is given by
\begin{equation}
v_{\text{eff}}=\sqrt{v(a)^2+v_{k}(a)^2}.
\label{eq_v_eff}
\end{equation}

We estimate the radio emission flux and characteristic frequency of 1027 exoplanets in our data set using the RBL (equations~\ref{eq_flux} and~\ref{eq_freq}). Table~\ref{tab1} lists the planetary, host star, and estimated radio emission parameters for these exoplanets. The complete version of this table is available in machine-readable format in the online journal.

\subsection{Key parameters for predicting radio emission}\label{key}
Large comparative studies of predicted radio emission properties across the known exoplanet population provide an opportunity to identify the key physical parameters associated with bright and potentially detectable radio emission, thereby improving our understanding of the conditions that favour strong exoplanetary radio sources \citep{2018MNRAS.478.1763L}. In general, applying the RBL framework to the entire population of known exoplanets is not feasible, as it requires a combination of planetary, stellar, and orbital parameters that are unavailable for a large fraction of discovered systems. Moreover, even for systems with complete parameter sets, the relative contribution of individual physical properties to the predicted radio emission remains difficult to quantify due to the complex and non-linear dependence of radio emission models on multiple parameters. Therefore, a data-driven approach capable of identifying the dominant physical parameters, revealing hidden relationships within the parameter space, and providing reliable predictions using a reduced set of input parameters would be valuable \citep{hinton1990connectionist}.

We apply two machine-learning interpretability techniques, Permutation Importance (PI) and SHapley Additive exPlanations (SHAP), to quantify the contribution of individual parameters to the model predictions obtained with the Parker-wind treatment and identify the dominant physical properties associated with exoplanetary radio emission. The PI method is a model inspection technique used to evaluate the contribution of each feature to a model's predictive performance (in this case, the random forest). It estimates feature importance by randomly shuffling the values of a feature and measuring the resulting degradation in model accuracy. A larger drop in accuracy indicates a greater importance of the feature in prediction \citep{2001MachL..45....5B}. Complementary to this approach, SHAP assigns each feature a contribution value to the prediction, based on Shapley values from cooperative game theory. It quantifies how much each feature contributes to pushing a prediction away from the average prediction. Unlike the PI, SHAP provides local interpretability, showing how individual features influence specific predictions, as well as global interpretability across the entire data set \citep{2017arXiv170507874L}.

\begin{figure*}
    \centering
    \includegraphics[width=0.49\textwidth]{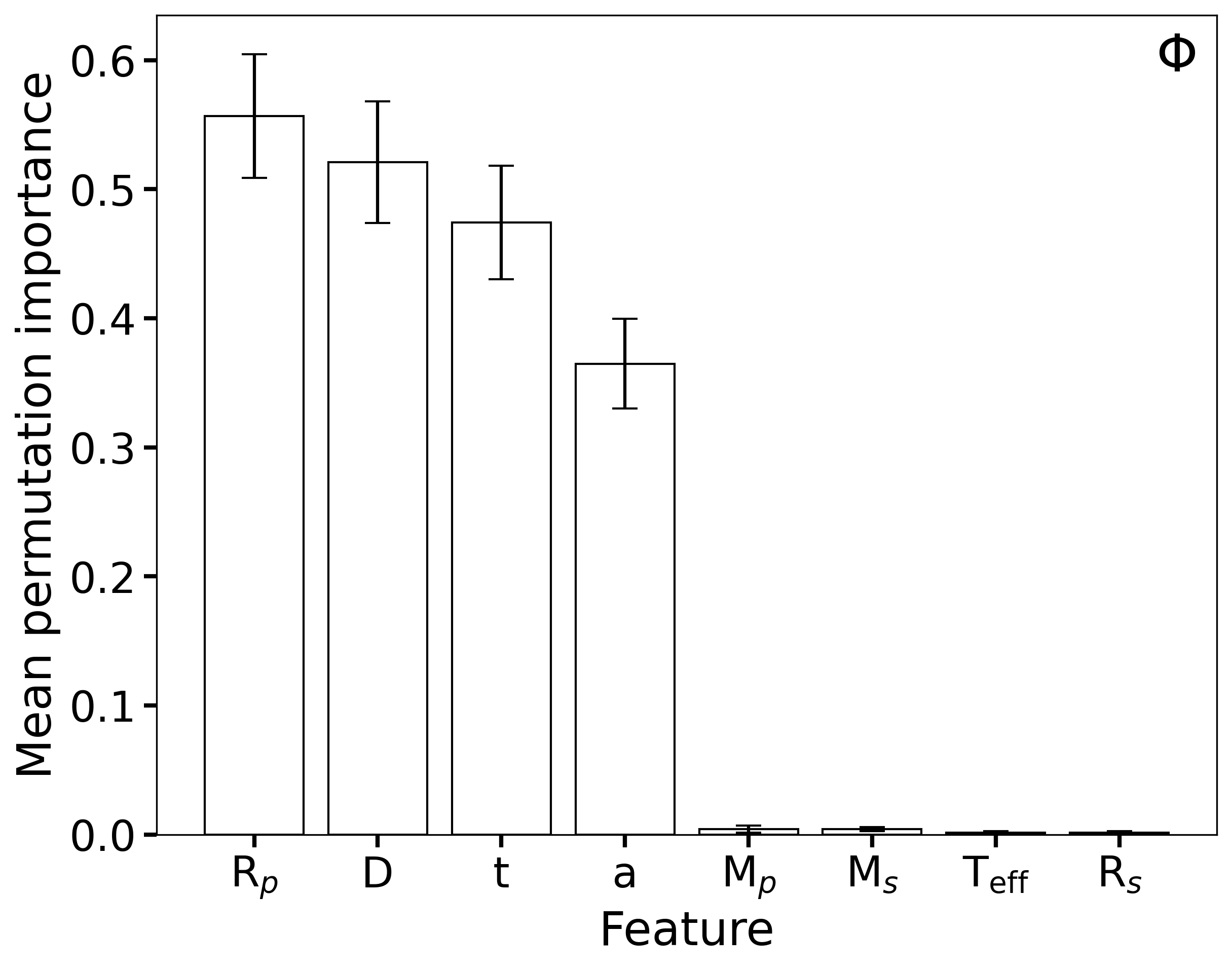}
    \includegraphics[width=0.49\textwidth]{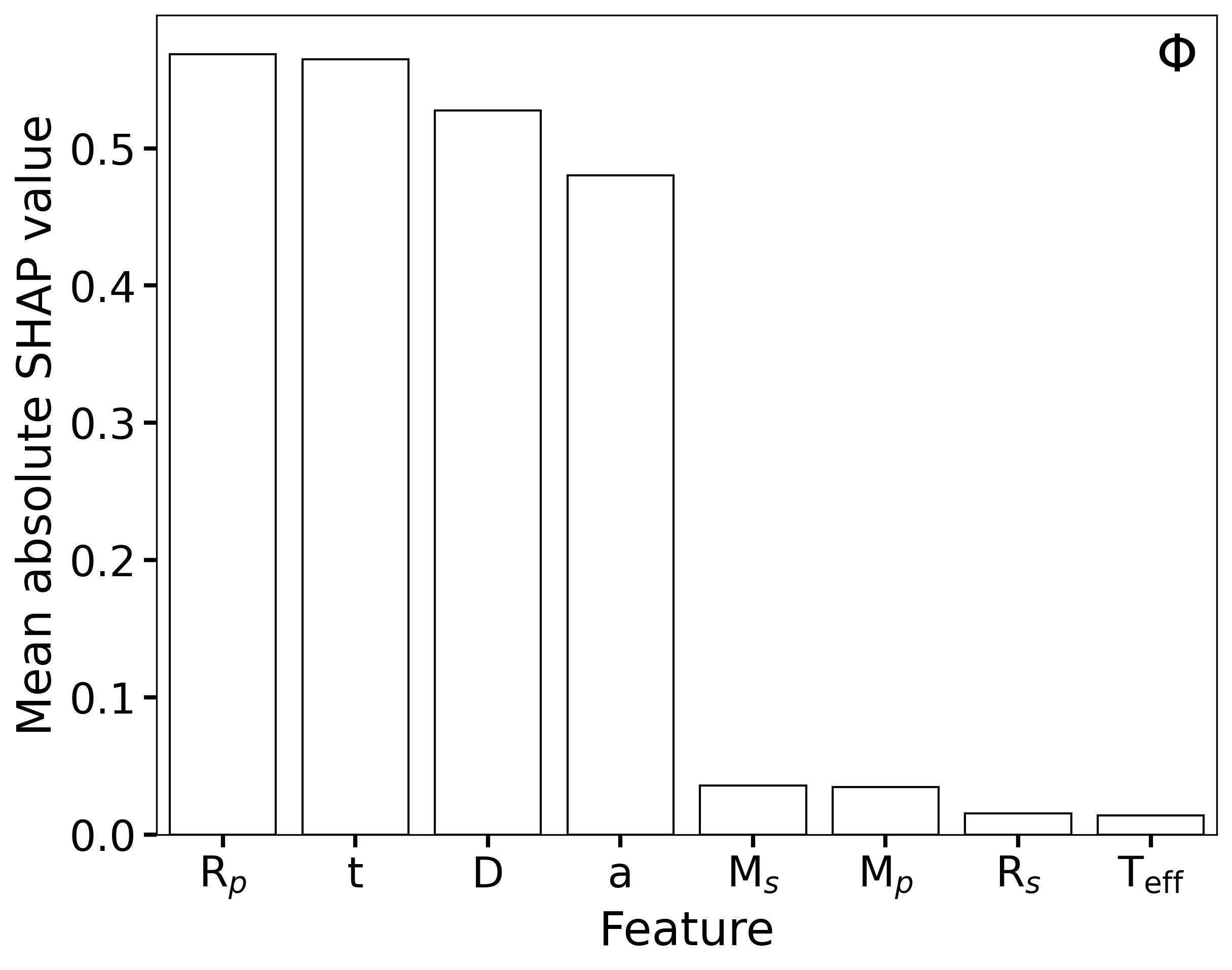}
    \includegraphics[width=0.49\textwidth]{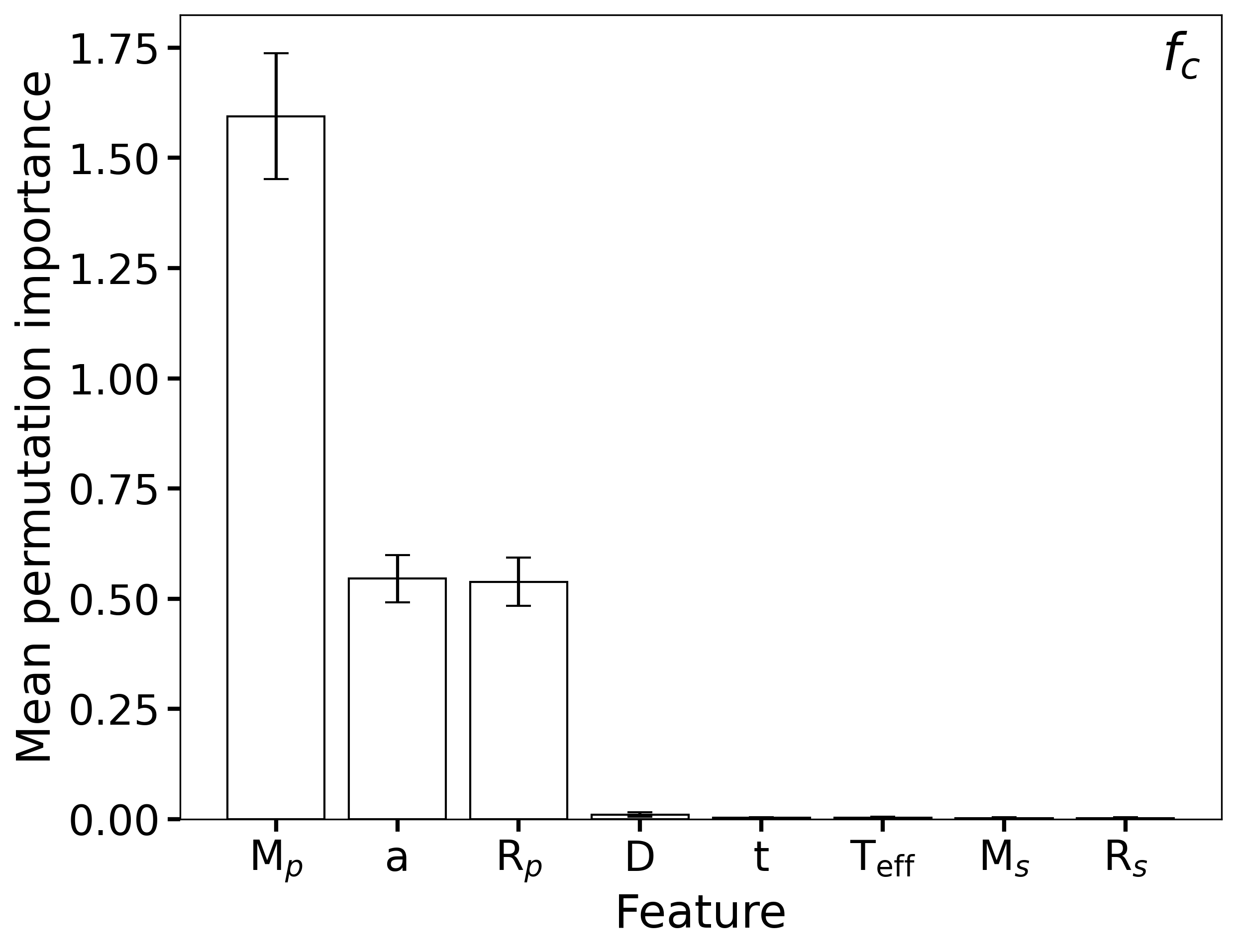}
    \includegraphics[width=0.49\textwidth]{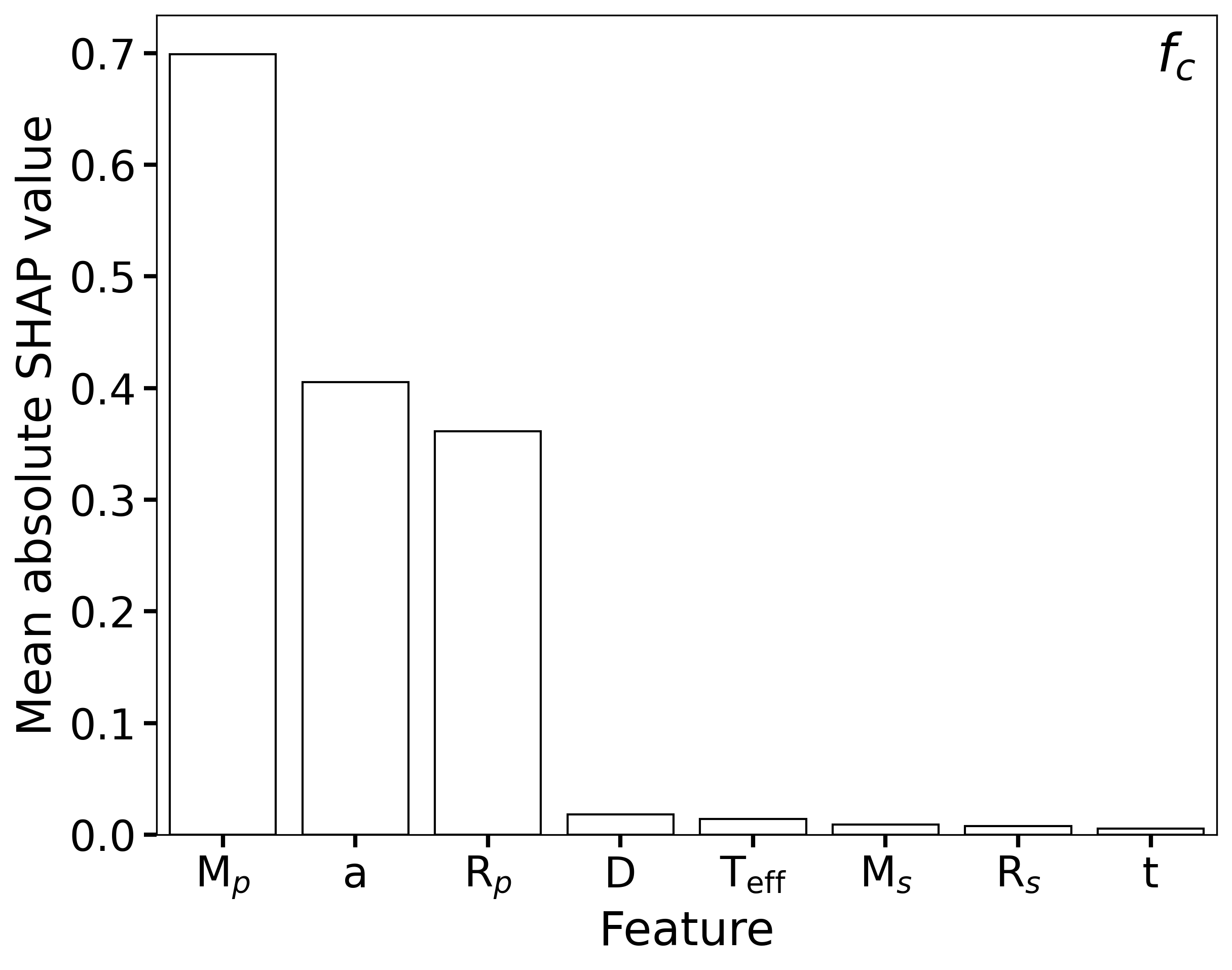}
    \caption{Feature-importance analysis for predicting exoplanetary radio flux (top panels) and characteristic frequency (bottom panels), showing mean permutation importance (left panels) and mean absolute SHAP values (right panels). The error bars represent the standard deviation of the permutation importance across 1000 random shuffles of each feature. For radio flux, the most influential features are planetary radius ($R_{p}$), Earth-star distance ($D$), stellar age ($t$), and semi-major axis ($a$), whereas for characteristic frequency, planetary mass ($M_{p}$) is the most important, followed by orbital semi-major axis ($a$) and planetary radius ($R_{p}$).}
    \label{fig_fs}
\end{figure*}

The eight fundamental physical parameters of exoplanetary systems, namely $M_p$, $R_p$, $M_s$, $R_s$, $T_{\mathrm{eff}}$, $t$, $a$, and $D$, are used as input features in the sensitivity analysis. Figure~\ref{fig_fs} compares the importance of these parameters. The upper left-hand panel shows the mean PI of the parameters used to predict the radio flux, where $R_{p}$ stands out with the highest importance of 0.557, followed by $D$ at 0.521, $t$ at 0.474, and $a$ at 0.365. The planet's mass and the host star's mass, radius, and effective temperature exhibit much lower feature importance values, each contributing less than $\sim0.05$. Similarly, the upper right-hand panel illustrates the mean absolute SHAP values, with $R_{p}$ (0.568) being the most influential, followed by $t$ (0.565), $D$ (0.528), $a$ (0.480), $M_{s}$ (0.036), $M_{p}$ (0.034), $R_{s}$ (0.016), , and $T_{\text{eff}}$ (0.014).

In our analysis, the feature-importance rankings are consistent with the physical structure of the revised model, while also capturing the effects of the empirical data distribution and feature correlations. For the flux density, they broadly follow the relative exponents in equation~\ref{eq_flux}, demonstrating that the machine learning model reproduces the expected physical scaling behaviour encoded in the equation. In particular, the dominant role of planetary radius arises directly from its exponent in the same relation. Despite the fact that the Earth-star distance enters through a clear geometric dependence, its exponent of -2 is smaller in magnitude than that of the planetary radius (+3), leading to a comparatively weaker dependence of the flux density on $D$. This statistical behavior has a clear physical interpretation within the RBL framework, where $R_{p}$ sets the effective magnetospheric cross-section and thus governs the efficiency of stellar wind energy interception and conversion into radio emission, making it a dominant factor in the intrinsic emitted power.

Although stellar age does not appear explicitly in the flux expression, it emerges as one of the dominant parameters because, through the adopted Parker-wind model, it influences the wind velocity and number density at the planet's orbit. These quantities determine the incident stellar-wind power and consequently affect the available energy input for auroral radio emission. Age therefore acts as an indirect driver of the flux prediction, entering through the wind model rather than the RBL formulation itself. Similarly, the semi-major axis is a dominant parameter because it defines the orbital location at which the stellar-wind properties are evaluated. Since the wind velocity and density vary strongly with distance from the star, $a$ directly regulates the stellar-wind ram pressure experienced by the planetary magnetosphere and hence the energy available for planetary radio emission.

Planetary mass, stellar mass, radius, and effective temperature, by contrast, show lower contributions to the flux prediction relative to the dominant parameters ($R_p$, $t$, $D$, and $a$). Planetary mass appears explicitly in the RBL magnetic-moment scaling but contributes negligibly to the flux prediction. This most likely results from its strong correlation with planetary radius in the working sample: Once $R_p$ is included as an input feature, $M_p$ provides little additional independent information, as the two parameters capture much of the same underlying variation in planetary structure. The correlation between $M_p$ and $R_p$ may lead to a redistribution of importance between these parameters in both PI and SHAP analyses. When features are correlated, SHAP can share importance between them, and PI may assign lower importance to a parameter if another correlated parameter carries similar information \citep{nicodemus2010behaviour,2023arXiv230305981V}. The sensitivity analysis should therefore not be interpreted as implying that planetary mass is physically irrelevant to the radio emission mechanism, but rather that, within the present sample and model formulation, its predictive influence is largely redundant with that of planetary radius. Consequently, removing $M_p$ from the parameter set does not measurably degrade the flux predictions (see appendix~\ref{rf}).

The mean PI of the parameters used to predict the characteristic frequency is shown in the lower left-hand panel of figure~\ref{fig_fs}. Among them, $M_{p}$ has the highest importance (1.595), followed by $a$ (0.546) and $R_{p}$ (0.538). The remaining parameters contribute minimally, each accounting for less than $\sim$0.01. The mean absolute SHAP values for frequency predictions are presented in the lower right-hand panel, where, similar to PI method, $M_{p}$ has the highest value (0.699), followed by $R_{p}$ (0.405) and $a$ (0.361). The remaining parameters have comparatively low importance, with individual contributions below $\sim$0.02.

The high importance of planetary mass and radius in predicting the emission frequency arises because the characteristic frequency is directly proportional to the planetary magnetic field strength, which itself depends on the planet's internal structure and dynamo properties. For the characteristic frequency, equation~\ref{eq_freq} indicates that $f_{c}\propto M_{p}^{5/3}R_{p}^{-3}$, implying a stronger dependence on planetary radius. This suggests that $R_{p}$ might play a more important role in predicting $f_{c}$; however, both PI and SHAP assign greater importance to $M_{p}$. This discrepancy likely arises from the properties of the data distribution. As shown in figure~\ref{fig1}, the dynamic range of planetary masses is significantly broader than that of planetary radii: the planetary mass range extends from $10^{-3}$ to 10$M_{J}$, while planetary radii span only 0.04-2$R_{J}$. Consequently, the variance in $R_{p}$ is much smaller than that in $M_{p}$. As feature-importance methods in machine learning are sensitive to the dynamic range and variance of input features, both PI and SHAP identify $M_{p}$ as the dominant contributor to variations in the predicted values of $f_{c}$, despite the larger exponent associated with $R_{p}$ in equation~\ref{eq_freq}. By contrast, equation~\ref{eq_flux} shows that the radio flux scales as $\Phi\propto M_{p}^{-0.33}R_{p}^{3}$, indicating a substantially stronger dependence on planetary radius than on mass. Unlike the frequency scaling (equation~\ref{eq_freq}), where the exponents are more comparable, the much larger exponent associated with $R_{p}$ (3 compared to -0.33 for $M_{p}$) dominates over the broader dynamic range of $M_{p}$, resulting in a higher feature importance for planetary radius in the flux predictions.

\begin{figure*}
    \centering
    \includegraphics[width=2.0\columnwidth]{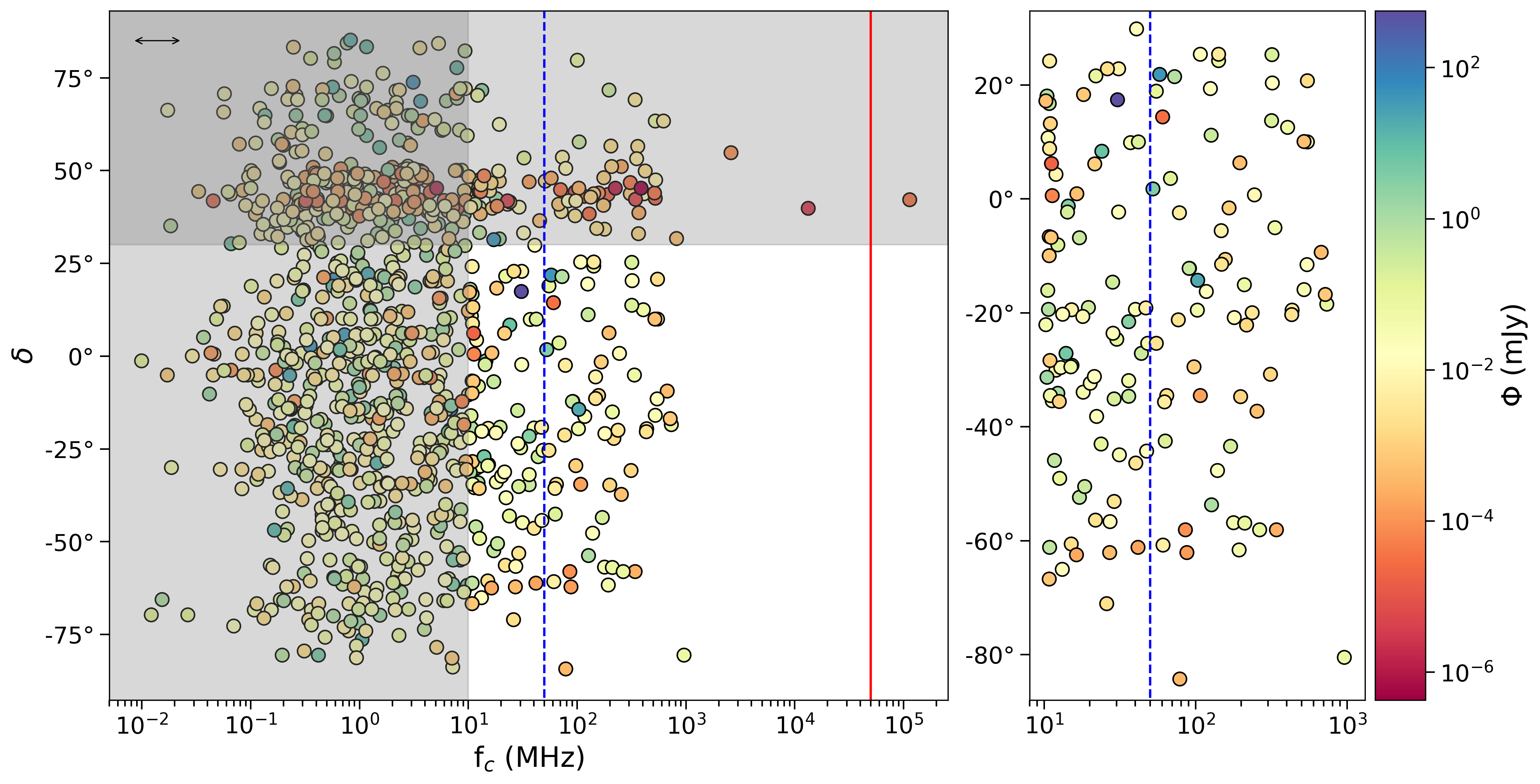}
    \caption{The predicted radio flux ($\Phi$) and characteristic frequency ($f_{c}$) of 1027 exoplanets (left panel). There are two grey-shaded areas: one indicating declinations ($\delta$) above +30$\degree$, which are not observable because of the telescope's geographic location and sky coverage, and the other representing emission frequencies below 10 MHz, which are unobservable from the ground due to the ionospheric cutoff. A total of 146 exoplanets lie outside these limiting areas (right panel). The blue dashed vertical line indicates the lower frequency boundary, while the red solid line marks the upper frequency boundary of the full SKA observing band.}
    \label{fig0}
\end{figure*}

Feature-importance analysis indicates that the semi-major axis also contributes to the prediction of the characteristic frequency. Although equation~\ref{eq_freq} does not explicitly depend on $a$, its effect may enter indirectly through the planetary rotation rate $\omega$. As described in section~\ref{rbl}, planets in close-in orbits ($a<0.1$ AU) are subject to strong tidal interactions that can result in tidal locking, a state expected for a significant fraction of such planets in our data set. Given that the planetary rotation rate enters equation~\ref{eq_freq}, the semi-major axis may indirectly affect the characteristic frequency through its role in tidal synchronization and rotation dynamics.

The interpretation of the role of $a$ corresponds to a more advanced physical description beyond the RBL formulation adopted here. The semi-major axis may influence the predicted characteristic frequency through its effect on the ionisation state and plasma environment of the planetary atmosphere. In addition to the intrinsic characteristic frequency, which is primarily determined by planetary properties such as $M_{p}$ and $R_{p}$, there exists an effective characteristic frequency. As discussed in section~\ref{quenched}, this effective frequency depends on whether CMI emission can successfully escape the planetary environment. In this context, the escape conditions may depend on the ionisation state and plasma environment of the planetary atmosphere, which in turn can be influenced by the orbital distance $a$ through stellar irradiation and stellar wind conditions. A complete treatment of this effect would also involve stellar parameters such as $T_{\mathrm{eff}}$, although these dependencies are not explicitly included in the current RBL formulation.

\begin{figure}
    \centering
    \includegraphics[width=1\columnwidth]{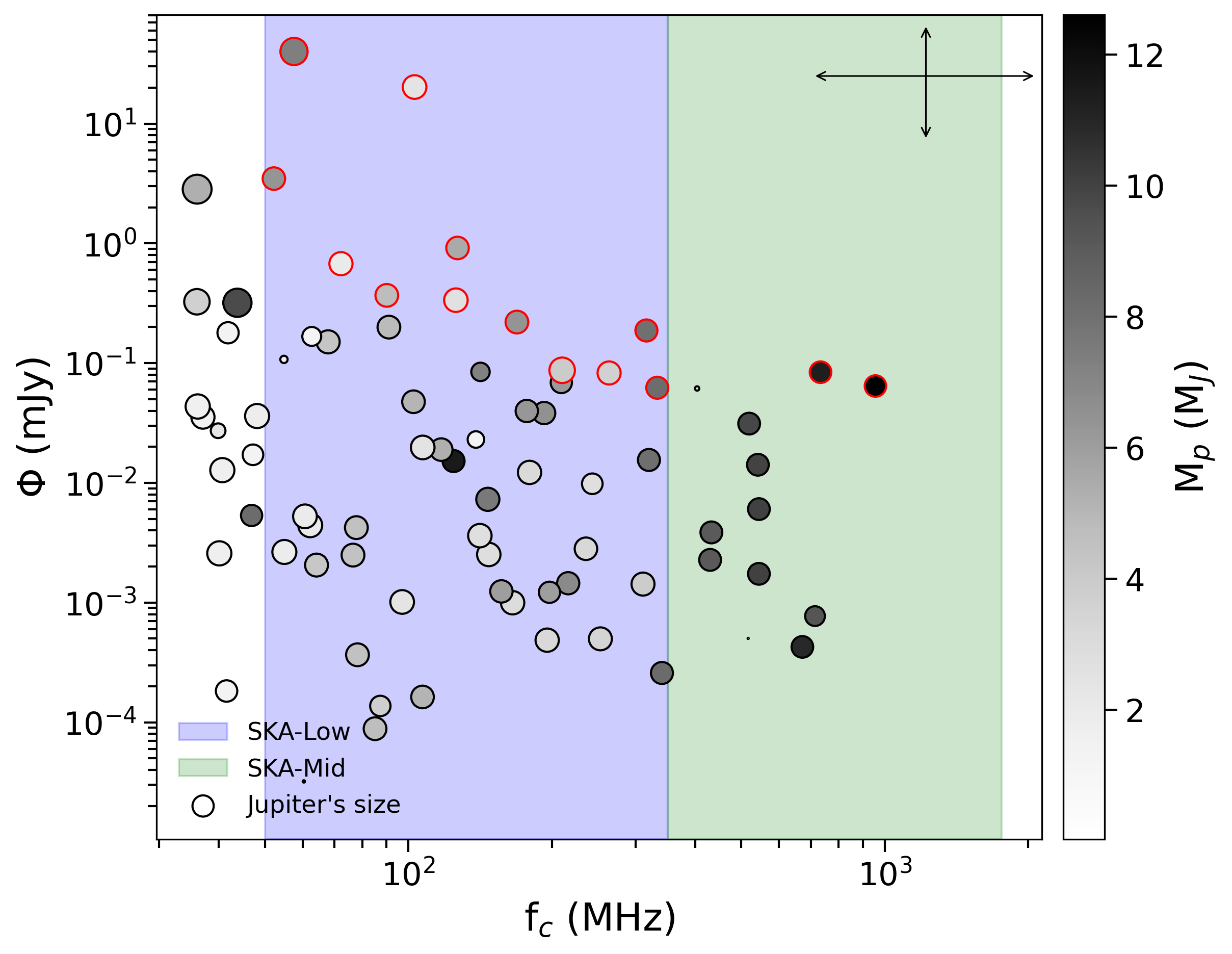}
    \caption{The predicted radio flux ($\Phi$) and characteristic frequency ($f_{c}$) of exoplanets with declinations below +30$\degree$, colour-coded by planetary mass ($M_{p}$). The symbol sizes have been scaled according to Jupiter's radius. The blue and green shaded areas indicate the frequency bands covered by SKA-Low and SKA-Mid, respectively. A total of 53 planets lie within the blue region, and 12 fall within the green region. Data points with red edges indicate planets predicted to be detectable with the SKA. Typical uncertainties for $\Phi$ and $f_{c}$ are shown by the arrows in the lower-right corner. For visual clarity, planets with $f_{c}<35$ MHz are not shown.}
    \label{fig5}
\end{figure}

Taken together, these results indicate that a reduced set of four parameters, including planetary radius, semi-major axis, stellar age, and distance, is sufficient to predict the radio flux density with reliable precision, while planetary mass is not required as an input feature. This is notable because planetary mass remains a challenging parameter to measure for a large fraction of exoplanets, often requiring dedicated radial velocity or transit timing follow-up observations. In principle, this approach provides a pathway toward estimating radio flux densities for the substantial population of known exoplanets that currently lack measured masses. However, this does not represent a complete assessment of radio detectability, as the characteristic emission frequency retains an explicit dependence on planetary mass. Therefore, although the flux density of a mass-incomplete planet can be estimated, the corresponding emission frequency, and hence whether the emission falls within an observable frequency range, cannot be reliably determined. Such flux predictions could nevertheless help prioritise future radial velocity follow-up observations by identifying planets with potentially strong radio emission but currently unconstrained masses. A subsequent mass measurement would then enable a complete assessment of their emission frequency and detectability within the full RBL framework. The methodology and predictive performance of a reduced-parameter surrogate model are described in appendix~\ref{rf}.

\section{Emissions detectable by SKA}\label{ska}
The RBL is used to estimate the radio flux ($\Phi$) and characteristic emission frequency ($f_{c}$) of 1027 exoplanets. However, ground-based detection of all these exoplanets is not feasible, as their emission frequencies might lie below the Earth's ionospheric cutoff. The Earth's ionosphere acts as a reflective barrier for low-frequency radio waves, effectively blocking transmissions below $\sim$10 MHz. This phenomenon, known as the ionospheric cutoff, prevents ground-based radio telescopes from detecting signals at these frequencies \citep{2007P&SS...55..598Z}. As a result, there are 794 planets in our sample with $f_{c}<$10 MHz, making them undetectable from the ground. Moreover, the sky coverage of a telescope plays an important role in determining the observability of an object. The SKA-Low stations are located at latitudes near -27$\degree$, while the SKA-Mid dishes are located around -31$\degree$ latitude \citep{2022JATIS...8a1021S,2022JATIS...8a1024L}. So, theoretically, the SKA can observe up to around +60$\degree$ in declination ($\delta$). However, in practice, telescopes generally observe targets above $30^{\circ}$ elevation to reduce atmospheric effects and optimise the signal-to-noise ratio (SNR). Accordingly, we adopt a more realistic declination limit of approximately $+30^{\circ}$ for the SKA, which results in the exclusion of an additional 87 exoplanets from the sample. The estimated radio flux for all 1027 exoplanets, along with their emission frequency and declination, is shown in figure~\ref{fig0}, where two grey-shaded areas represent declinations above +30$\degree$ and emission frequencies below 10~MHz. As shown, 146 exoplanets of the sample have $f_{c}$ greater than 10~MHz and are located within regions of the sky observable by the SKA Observatory.

SKA-Low operates within a single frequency band, covering the range from 50 to 350 MHz. SKA-Mid's full frequency band extends up to 50 GHz; however, the ranges between 1.76-4.6 GHz and above 15.4 GHz are not included in the current deployment plan. Figure~\ref{fig5} presents the calculated radio flux of exoplanets with $\delta<+30\degree$, plotted as a function of maximum emission frequency. In this figure, the colour and size of the symbols represent the planetary mass and radius, respectively, relative to those of Jupiter. The frequencies covered by SKA-Low and SKA-Mid are illustrated by blue and green shaded areas, respectively. Consequently, a total of 53 exoplanets have maximum emission frequencies within the SKA-Low band, and 12 fall within the SKA-Mid frequency bands.

\begin{table}
\centering
\caption{The 1$\sigma$ imaging sensitivities for continuum observations with SKA \citep{2019arXiv191212699B}. The first column shows the row number. Columns 2-4 present the minimum, central, and maximum observing frequencies, while columns 5 and 6 give the corresponding imaging sensitivities for the AA4 and AA* subarray configurations, respectively. Rows 1-6 correspond to SKA-Low, while rows 7-15 correspond to SKA-Mid.}
\begin{tabular}{lllllll}
\hline
\# & \makecell[l]{$\nu_{\min}$\\(MHz)} & \makecell[l]{$\nu_{c}$\\(MHz)} & \makecell[l]{$\nu_{\max}$\\(MHz)} & \makecell[l]{$\sigma_{\mathrm{AA4}}$\\($\mu$Jy/B)} & \makecell[l]{$\sigma_{\mathrm{AA*}}$\\($\mu$Jy/B)}\\
\hline
1  & 50    & 60    & 69    & 163  & 436  \\
2  & 69    & 82    & 96    & 47   & 126  \\
3  & 96    & 114   & 132   & 26   & 70   \\
4  & 132   & 158   & 183   & 18   & 48   \\
5  & 183   & 218   & 253   & 14   & 37   \\
6  & 253   & 302   & 350   & 11   & 29   \\
7  & 350   & 410   & 480   & 16.8 & 26.0 \\
8  & 480   & 560   & 650   & 8.1  & 11.1 \\
9  & 650   & 770   & 890   & 4.4  & 6.0  \\
10 & 890   & 1050  & 1210  & 2.7  & 3.7  \\
11 & 1210  & 1430  & 1650  & 2.0  & 2.7  \\
12 & 4180  & 4940  & 5700  & 1.4  & 2.2  \\
13 & 5700  & 6740  & 7780  & 1.3  & 2.0  \\
14 & 7780  & 9190  & 10610 & 1.2  & 1.9  \\
15 & 10610 & 12530 & 14460 & 1.2  & 1.9  \\
\hline
\end{tabular}
\label{tab2}
\end{table}

\begin{table*}
\centering
\caption{List of exoplanets predicted to be detectable with SKA. Columns 1 and 2 present the row number and the planet's name. Columns 3 and 4 lists the maximum emission frequency and flux density of the planets, respectively. Columns 5 and 6 indicate the telescope and the corresponding subarray configuration under which the planet is expected to be detectable, based on the 5$\sigma$ imaging sensitivity of the AA4 and AA* for continuum observations, assuming a bandwidth of 30\% of the central observing frequency and an integration time of 1 hour. The table has been sorted by flux density in descending order.}
\begin{threeparttable}
\begin{tabular}{llllllllllllll}
\hline
\# & Name & \makecell[l]{$f_c$\\(MHz)} & \makecell[l]{$\Phi$\\(mJy)} & Telescope & \makecell[l]{Subarray\\configuration} \\
\hline
1  & WASP-14 Ab     & 57.5   & 40.064 & SKA-Low & AA4, AA* \\
2  & GJ 876 b$^{\S}$       & 103.1  & 20.219 & SKA-Low & AA4, AA* \\
3  & TOI-628 b      & 52.2   & 3.477  & SKA-Low & AA4, AA* \\
4  & HD 89839 b$^{\S}$     & 126.8  & 0.915  & SKA-Low & AA4, AA* \\
5  & TOI-4515 b$^{\dagger}$     & 72.2   & 0.677  & SKA-Low & AA4, AA* \\
6  & HD 82943 c$^{\S}$     & 90.1   & 0.368  & SKA-Low & AA4 \\
7  & HD 196885 Ab$^{\S}$   & 125.8  & 0.335  & SKA-Low & AA4 \\
8  & TOI-2005 b     & 169.0  & 0.220  & SKA-Low & AA4 \\
9  & 70 Vir Ab$^{\S}$      & 316.0  & 0.187  & SKA-Low & AA4, AA* \\
10 & TOI-4641 b     & 317.5  & 0.185  & SKA-Low & AA4, AA* \\
11 & TOI-3362 b     & 210.3  & 0.087  & SKA-Low & AA4 \\
12 & HD 141937 b$^{\S}$    & 733.1  & 0.084  & SKA-Mid & AA4, AA* \\
13 & TOI-558 b      & 263.9  & 0.083  & SKA-Low & AA4 \\
14 & $\pi$ Men b$^{\S}$    & 955.9  & 0.064  & SKA-Mid & AA4, AA* \\
15 & HD 7449 Ab$^{\S}$     & 333.0  & 0.062  & SKA-Low & AA4 \\
\hline
\end{tabular}
\begin{tablenotes}
\item$^{\dagger}$The planet might experience radio quenching due to $f_{p}>f_{c}$ (see section~\ref{quenched} for further details).
\item$^{\S}$The planetary radius has been derived using equation~\ref{eq1}, which may introduce additional uncertainties into the predicted radio emission properties (see section~\ref{uncr}).
\end{tablenotes}
\end{threeparttable}
\label{tab3}
\end{table*}

To determine whether the predicted radio emission of an exoplanet is detectable with the SKA, it is essential to consider the imaging sensitivity of both SKA-Low and SKA-Mid for continuum observations. As discussed by \cite{2019arXiv191212699B}, the image noise is given by
\begin{equation}
\sigma=\frac{S_{D}\;SEFD}{\eta_{s}\sqrt{n_{\text{pol}}\Delta\nu\Delta\tau}},
\label{eq13}
\end{equation}
where $S_{D}$ represents a degradation factor relative to the natural array sensitivity. $\eta_{s}$ denotes the system efficiency, which includes the finite correlator efficiency and other sources of incoherence, and is typically assumed to be $\eta_{s}=0.9$. The factor $n_{\text{pol}}$ corresponds to the number of contributing polarisations, which for the SKA, equipped with orthogonal linear polarisations, is $n_{\text{pol}}=2$. The contributing bandwidth and integration time are denoted by $\Delta\nu$ and $\Delta\tau$, respectively. The System Equivalent Flux Density (SEFD) is defined by
\begin{equation}
SEFD=\frac{2k_{B}T_{\text{sys}}}{A_{\text{eff}}},
\label{eq14}
\end{equation}
where $k_{B}$ is the Boltzmann constant, $T_{\text{sys}}$ the system temperature, and $A_{\text{eff}}$ the effective collecting area. Together, these factors determine the effective sensitivity of the array for a given observation, providing a realistic estimate of its capability to detect faint radio sources.

For SKA observations, we consider the SKA-Low frequency range, spanning 50-350 MHz, and the SKA-Mid frequency ranges, covering 0.35-1.76 GHz (bands 1 and 2) and 4.6-15.4 GHz (band 5). Assuming continuum observations with a fractional bandwidth of approximately 30\% of the central frequency ($\Delta\nu/\nu_{c}=0.3$, where $\nu_{c}$ is the central observing frequency) and an integration time of $\Delta\tau=1$~hour, the corresponding $1\sigma$ imaging sensitivities for the AA4 and AA* subarray configurations are summarised in table~\ref{tab2}. These values offer a realistic estimate of the array's observational capabilities and serve as a reference for determining which planets are likely detectable with the SKA. For a more conservative evaluation of exoplanetary radio detectability, we adopt the 5$\sigma$ imaging sensitivities for SKA-Low and SKA-Mid.

Table~\ref{tab3} presents the predicted radio emission parameters for the exoplanets potentially detectable with SKA. The last column of this table indicates the subarray configuration with which each planet is expected to be detectable. As a result, the AA4 configuration of SKA-Low is anticipated to detect 13 exoplanets, whereas the more compact AA* configuration is expected to recover 7 of these 13 sources. SKA-Mid, in its AA4 stage, can potentially detect two exoplanets, both of which remain detectable with the AA* configuration. Among these SKA targets, the most promising candidates are those with the highest predicted flux densities within the SKA-Low frequency range, while the SKA-Mid targets are distinguished by their higher characteristic emission frequencies and correspondingly favourable flux densities. In particular, WASP-14 Ab (40.064 mJy at 57.5 MHz) and GJ 876 b (20.219 mJy at 103.1 MHz) represent the most favourable SKA-Low candidates, whereas HD 141937 b (0.084 mJy at 733.1 MHz) and $\pi$ Men b (0.064 mJy at 955.9 MHz) are the principal targets for SKA-Mid. These sources could therefore represent some of the most promising candidates for future targeted radio observations with the SKA.

\section{Discussion}\label{disc}
In this section, we address radio-quenching effects in section~\ref{quenched}, discuss model assumptions and uncertainties associated with our framework in section~\ref{uncr}, address relevant observational considerations in section~\ref{obs_cons}, and compare our predictions with recent observational campaigns and existing radio surveys in section~\ref{vla}.

\subsection{Radio-quenched exoplanets}\label{quenched}
A planetary magnetic field is a prerequisite for generating the radio emission through the CMI, but its presence alone does not guarantee emission. For CMI to operate, the local electron cyclotron frequency ($f_{c}$) must exceed the surrounding plasma frequency ($f_{p}$) at the emission region, satisfying the condition $f_{p}/f_{c}<0.4$ \citep{1985AnGeo...3..273L,1992GeoRL..19..237H,2001P&SS...49.1137Z}.

For low-mass planets and those in extremely close-in orbits, intense X-ray and extreme ultraviolet flux from the host star heats the upper atmosphere, causing its expansion and increasing the plasma density within the magnetosphere. This expanded, ionised atmospheric gas may suppress or entirely quench the expected radio emission \citep{2017AN....338..881D,2017MNRAS.469.3505W,2018MNRAS.480.3680W,2018MNRAS.479.1194D,2022MNRAS.512.4869E}. However, for planets with higher masses, the planetary atmosphere remains hydrostatic even in close-in orbits, enabling sustained radio emission. Consequently, the plasma frequency becomes sensitive to planetary mass, orbital distance, and stellar properties. \citet{2023pre9.conf03090G} identified a region of parameter space, defined by planetary mass and orbital distance, where radio quenching becomes significant. They showed that for planets with orbital distances below 0.2\;AU and masses between 0.01 and 2\;$M_{J}$, an extended planetary atmosphere can quench the radio emission, whereas this effect is less pronounced in other regions of the parameter space.

\begin{figure}
    \centering
    \includegraphics[width=1\columnwidth]{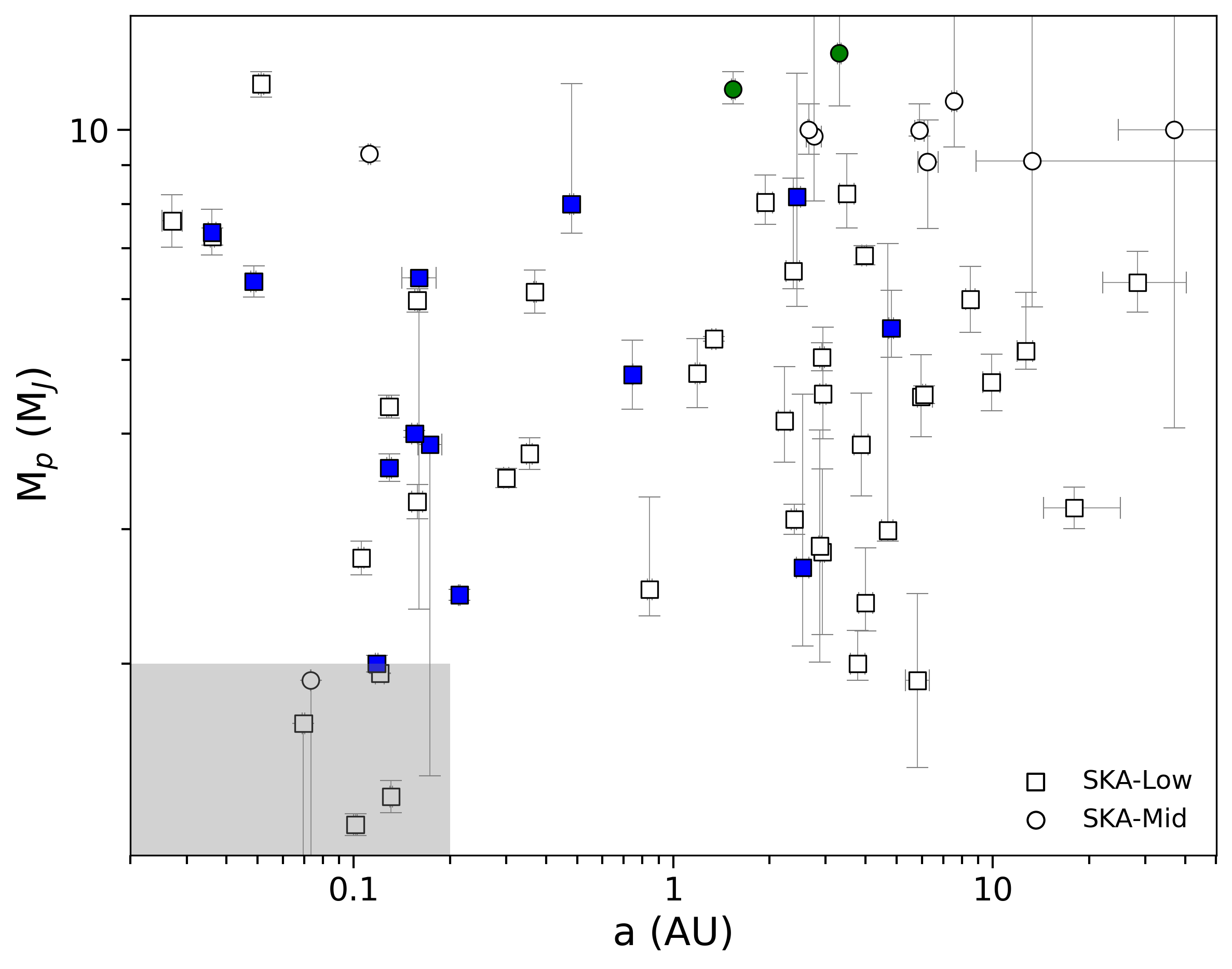}
    \caption{Planetary mass ($M_{p}$) as a function of orbital semi-major axis ($a$) for 53 planets with predicted emission frequencies falling within the SKA-Low band (squares) and 12 planets within the SKA-Mid band (circles). Blue filled squares mark planets expected to be detectable with SKA-Low, and green filled circles denote those expected to be detectable with SKA-Mid, according to the 5$\sigma$ imaging sensitivity for continuum observations with SKA. The grey-shaded region indicates the parameter space defined by \protect\cite{2023pre9.conf03090G}, with $a<0.2$\,AU and $0.01<M_{p}\leq2\,M_{J}$, where extended planetary atmosphere may lead to radio quenching. For visual clarity, planets HIP 41378 b and K2-184 b, with masses below 0.04~$M_J$, are not shown.}
    \label{fig_quench}
\end{figure}

Figure~\ref{fig_quench} presents planetary mass as a function of orbital semi-major axis for 65 exoplanets within the SKA-Low and SKA-Mid frequency bands. As shown, two planets in the SKA-Mid band fall within the radio-quenching region, while six planets with predicted emission frequencies in the SKA-Low band are located in this region. For SKA-Low, they are K2-184 b, WASP-130 Ab, HATS-17 b, K2-95 b, TOI-2179 b, and TOI-4515 b, while the two SKA-Mid targets are HIP 41378 b and K2-77 b. Notably, TOI-4515 b (marked in table~\ref{tab3}), is predicted to be detectable but lie within this region, indicating this is a suboptimal target for SKA-Low observations. This highlights the importance of accounting for radio quenching when selecting targets for observation campaigns such as those planned with the SKA Observatory.

\subsection{Model assumptions and uncertainties}\label{uncr}
\subsubsection{Uncertainties in planetary radii and RBL input assumptions}\label{uncr_radii}
A major source of uncertainty in our framework arises from planets lacking direct radius measurements, for which $R_{p}$ is estimated using the mass-radius relation (equation~\ref{eq1}). However, this relation provides only an approximate estimate of planetary radius and does not account for the substantial intrinsic scatter in exoplanet populations (see figure~\ref{fig1}, where planets with comparable masses display considerable variation in radius). Moreover, uncertainties in the measured planetary mass could propagate directly into the estimated radius. Consequently, the derived radius should not be interpreted as a precise value. To quantify the impact of this uncertainty, we propagate the upper and lower bounds of the mass-radius relation through the RBL for the 109 planets in our sample lacking directly measured radii. This yields an average logarithmic spread of $\sim0.94$ dex in both the predicted characteristic frequency and radio flux density, corresponding to nearly an order-of-magnitude uncertainty. This amplification arises from the strong dependence of both quantities on planetary radius, where $\Phi\propto R_{p}^{3}$ and $f_{c}\propto R_{p}^{-3}$ (see equations~\ref{eq_flux} and~\ref{eq_freq}).

This propagation of uncertainties is particularly important for detectability. For eight planets without directly measured radii that fall within the potentially detectable flux and frequency ranges, we adopt a conservative criterion requiring that the emission frequency remains within the instrumental bands and that the flux exceeds the sensitivity threshold across the full uncertainty range; planets that do not satisfy either condition are classified as marginal candidates. As a result, under the worst-case combination, all eight of these systems fail at least one of the detectability conditions. Specifically, GJ 876 b, HD 89839 b, HD 82943 c, HD 196885 Ab, 70 Vir Ab, and HD 7449 Ab have lower limits of their characteristic frequencies below the SKA-Low observing range ($>$50 MHz), while HD 141937 b and $\pi$ Men b have upper frequency bounds exceeding the upper limit of the SKA-Mid observing range (15.4 GHz).

Beyond the uncertainty in planetary radius estimates, further uncertainties may arise from the simplifying assumptions adopted for the planetary magnetic moment. To compute the radio power generated by an exoplanet, understanding the planet's magnetic field strength and its temporal evolution is required \citep{2009Natur.457..167C}. According to \citet{2010A&A...522A..13R}, the average magnetic field strength of giant planets and brown dwarfs can vary by approximately one order of magnitude or more over their lifetimes, exhibiting stronger fields during their youth when they are more luminous. Assuming an orbital separation of 0.1 AU, the magnetic field for hot Jupiters declines from 240 G at 500 Myr to 120 G at 5 Gyr, while for hot Neptunes, it starts at 11 G at young ages and diminishes at $\gtrsim$2 Gyr \citep{2024MNRAS.535.3646K}. Determining a planet's magnetic field strength over its lifetime demands intricate modelling and numerical simulations of its interior structure. To simplify this, we employ Blackett-type scaling laws, which relate a planet's magnetic moment to its rotation rate and mass \citep{1947Natur.159..658B}. We note, however, that this reliance on the classical Blackett-type prescription for estimating planetary magnetic moments is a limitation shared by the RBL framework. Theoretical studies of planetary dynamos suggest that the convective processes driving the magnetic field depend little on planetary rotation, and are instead primarily powered by thermal or compositional gradients within the planetary interior \citep{2003E&PSL.208....1S,2006GeoJI.166...97C,2010SSRv..152..651S,2011Icar..213...12D}. Consequently, the estimated magnetic field strengths and radio emission properties reported here, and in comparable prior studies, may differ when alternative dynamo prescriptions are adopted. Taken together, the simplified treatment of the planetary magnetic moment adopted here represents first-order approximations whose refinement, together with more precise planetary radius measurements, would improve the reliability of RBL-based radio emission predictions.

\subsubsection{Uncertainties in extrapolating the RBL to diverse exoplanet populations}
A further limitation of our framework concerns the applicability of the RBL across diverse planetary populations, rather than the precision of its individual inputs. The properties of a planet's auroral radio emission depend on various factors, such as the strength and topology of its magnetic field. The magnetic field itself depends critically on the planet's interior structure, composition, thermal evolution, and dynamo processes \citep{2018A&A...616A.182V,2019BAAS...51c.135L,2024arXiv240415429B}. A long-lived planetary magnetic field can be generated through dynamo process if the planet possesses an electrically conductive convecting fluid in its core and the mantle allows sufficiently rapid core cooling. The temperature at the core-mantle boundary is particularly important, as it largely controls the heat flux out of the core and thereby determines whether a dynamo can be sustained. Venus's core, for instance, may not be cooled sufficiently to maintain an active dynamo at present, possibly compounded by the absence of an inner core \citep{2003E&PSL.208....1S,2007cody.book...31N,2010SSRv..152..565C,2019AdPhX...430316V}.

Super-Earths and sub-Neptunes, planets of sizes ranging from 1 to 4 Earth radii, orbit at least 30\% of sun-like stars in the Milky Way and represent the most numerous class of known exoplanets \citep{2010Sci...330..653H,2013ApJS..204...24B,2014ApJ...790..146F,2025ApJ...979L..23S}. Super-Earths are generally interpreted as predominantly rocky planets, with interior structures broadly analogous to scaled-up terrestrial planets in the solar system. Their interiors are expected to comprise an iron-rich core surrounded by a silicate mantle \citep{2007ApJ...665.1413V,2007lyot.confE..39K}. In contrast, sub-Neptunes are interpreted as planets possessing substantial volatile-rich envelopes, whose interiors typically feature a rocky or icy core overlain by high-pressure ice or silicate mantle layers and surrounded by a thick hydrogen-helium atmosphere \citep{2014ApJ...792....1L,2019PNAS..116.9723Z,2021JGRE..12606639B}. Differences in the interior structures and compositions of super-Earths and sub-Neptunes, compared to solar system planets, are expected to produce markedly different planetary magnetic fields and, consequently, distinct characteristics in their auroral radio emissions. Massive super-Earths might sustain a dynamo in their metallic cores provided their silicate mantles remain fully solid, while sub-Neptunes, akin to Uranus and Neptune, may suffer suppressed core heat flux due to insulating interior layers, potentially leading to weaker or absent dynamos \citep{2020RSPTA.37890479S,2021AGUFMGP45E0446B,2024PNAS..12103981M,2025JGRE..13008550W}.

The RBL used in our study is a semi-empirical scaling relation calibrated on solar system planets with active dynamos and detected auroral radio emissions. It does not explicitly incorporate detailed interior dynamics, such as core stratification or mantle convection efficiency. Hence, the structural and compositional diversity of exoplanets could introduce considerable uncertainties when employing the RBL to estimate planetary radio emission, particularly for super-Earths and sub-Neptunes, which lack direct analogues in the solar system.

Moreover, the RBL may overestimate radio emission from Venus-like terrestrial exoplanets. Although Earth and Venus have nearly identical bulk properties, Earth sustains a strong global magnetic field with associated radio emission potential, whereas Venus lacks such a dynamo-generated field \citep{2014PEPI..236...36D}. The RBL implicitly assumes that planets with similar mass, rotation rate, orbital parameters, and stellar wind conditions generate comparable radio power. It does not account for cases where a dynamo is weak or absent, even when bulk properties resemble those of radio-emitting solar system planets. Therefore, it cannot distinguish between dynamo-active planets such as Earth and dynamo-inactive cases like Venus. Nevertheless, such planets in our sample are predicted to produce fluxes below SKA detection thresholds, leaving gas giants, with more reliable Jupiter-like analogues, as the most promising detectable candidates.

\subsection{Observational considerations}\label{obs_cons}
The detectability assessment presented in this work is based on the nominal 5$\sigma$ continuum sensitivities of the SKA, assuming a 1-hour integration time. However, the practical detectability of exoplanetary radio emission also depends on several observational considerations. Although the imaging sensitivity improves with increasing integration time as described by equation~\ref{eq13}, the likelihood of successful detection is expected to be limited by additional physical and observational factors. In particular, CMI emission is believed to be highly variable, strongly beamed, and modulated by planetary rotation and orbital geometry \citep{1999JGR...10414025F,2007P&SS...55..598Z}. The beaming properties of the emission, combined with the changing viewing geometry relative to Earth, can lead to substantial variations in the observed signal. In addition to orbital motion, this variability can arise from planetary rotation in non-tidally locked systems and from stellar rotation. Planetary radio emission is also expected to vary in response to changes in the stellar wind or interactions with stellar CMEs, further complicating the interpretation of any detected signals \citep{2010MNRAS.406..409F,2018haex.bookE.159G}. Consequently, simply increasing the integration time may not always improve the likelihood of detection. Instead, a more realistic observing strategy would involve repeated observations or scheduling them to coincide with favourable orbital or rotational phases of the planet. We further note that, once the noise floor approaches the confusion limit set by the dense low-frequency background source population, increasing the integration time no longer improves the sensitivity, thereby establishing a fundamental limit that is independent of $\Delta\tau$.

A further consideration for current and future radio observations, including those with the SKA, is the impact of the radio-frequency interference (RFI) environment. Although the SKA is designed with advanced RFI mitigation capabilities, residual human-made interference may remain a limiting factor, particularly for the detection of extremely weak exoplanetary radio signals that can lie below the noise level of raw observations. Continued improvements in calibration and RFI mitigation techniques are therefore essential to fully exploit the theoretical sensitivity of the SKA under realistic observing conditions.

Furthermore, the detectability thresholds adopted throughout this work are based solely on total-intensity (Stokes~I) sensitivity. Planetary CMI emission is expected to be strongly, and in some cases nearly completely, circularly polarised \citep{2025ARA&A..63..299V}. This conclusion follows simply from the nature of cyclotron emission itself, and circularly polarised emission is the standard for planetary radio emission in the solar system. Since most astrophysical background sources and instrumental leakage are only weakly circularly polarised, observations in Stokes~V can significantly enhance the effective SNR by substantially reducing contamination from unpolarised emission. Therefore, the detectability estimates presented here should be regarded as first-order predictions based primarily on instrumental sensitivity. More realistic observing strategies will require careful consideration of the temporal behaviour of the planetary radio emission, together with the substantial benefits offered by observations in circular polarisation.

\subsection{Comparison with previous observations}\label{vla}
Large-scale surveys have been done to detect and confirm the first radio signal from an exoplanet; however, no such detection has yet been confirmed. A comprehensive search for low-frequency radio emission from 175 exoplanetary systems was conducted by \cite{2014A&A...562A.108S} using data from the TIFR GMRT Sky Survey (TGSS) at 150 MHz, supplemented by searches in the NRAO VLA Sky Survey (NVSS) and VLA FIRST survey at 1.4 GHz \citep{1998AJ....115.1693C}. For 61 Vir, HD 86226, HD 164509, and 1RXS 1609, TGSS radio sources were identified, coinciding with, or located very close to their sky positions. Radio emission toward the planet-hosting pulsar PSR B1620-26 and the giant planet HD 43197 b was reported in the NVSS, while emission coincident with the positions of the hot Jupiter WASP-77 A b was detected in the FIRST survey.

For 61 Vir, the TGSS source reported by \cite{2014A&A...562A.108S} was subsequently resolved in higher-resolution VLA observations by \cite{2021MNRAS.500..211G} and identified as an extragalactic double source significantly offset from the stellar position. This is consistent with our predictions: 61 Vir b and c are included in our analysed sample (see rows 10 and 11 of table~\ref{tab1}), and we predict both planets to be effectively unobservable from the ground, with characteristic emission frequencies of 0.5 and 2.2~MHz, respectively, both well below the Earth's ionospheric cutoff of $\sim$10~MHz. The remaining candidate sources toward HD 86226, HD 164509, 1RXS 1609, PSR B1620-26, HD 43197, and WASP-77 A are not included in our analysis because they do not satisfy the selection criteria adopted for constructing our sample.

We identify HD 141937 b as a potentially detectable SKA-Mid target, with a predicted flux density of 0.084~mJy at 733.1~MHz (see row 12 of table~\ref{tab3}). However, no radio emission was detected from HD 141937 by \cite{2014A&A...562A.108S} in their GMRT observations, possibly due to factors such as emission not directed toward the observer, a highly variable radio emission, an unmagnetised planet, a weaker-than-expected stellar wind, or insufficient instrument sensitivity \citep{2007P&SS...55..598Z}.

In a highly sensitive GHz-frequency survey with VLA, \cite{2024AJ....168..127O} investigated 77 stellar systems hosting 140 exoplanets, primarily within 17.5 pc, spanning masses from 0.001 to 10 $M_{J}$ and semi-major axes from 0.01 to 10 AU. Radio emission was detected from only one target, GJ 3323, which exhibited a circular polarisation fraction of approximately 40\%. The detected radio luminosity corresponds with host star's known X-ray luminosity and the Güdel-Benz relation for stellar activity, indicating a likely stellar origin; however, the high circular polarisation fraction may also suggest star-planet interactions. GJ 3323 has been excluded from our predictions due to the absence of a complete set of parameters necessary for calculating the radio emission properties. There are 13 systems in our sample, comprising 23 planets, that were targeted by \citet{2024AJ....168..127O}: GJ 876, GJ 581, 55 Cnc, GJ 436, GJ 1214, HD 260655, LHS 1140, $\rho$ CrB, HD 69830, HD 87883, 51 Peg, $\tau$ Boo, and 70 Vir. All observations resulted in non-detections. For the two planets that we identify as potentially detectable with the SKA-Low, GJ 876 b and 70 Vir Ab (see rows 2 and 9 of table~\ref{tab3}), we predict characteristic emission frequencies of 103.1~MHz and 316.0~MHz, respectively, which are approximately 13-75 times lower than the frequencies probed by the observations. As CMI emission is expected to have a sharp high-frequency cutoff near its characteristic emission frequency rather than extending substantially to higher frequencies, the GHz non-detections do not place meaningful constraints on the predicted MHz-frequency emission. The GJ 876 planetary system was also identified as capable of generating radio emission above 10 $\mu$Jy by \citet{2018ApJ...854...72T}.

\cite{2025A&A...700A.140Z} conducted NenuFAR (New extension in Nançay upgrading LOFAR) imaging observations of HD 189733 system in the 15-62 MHz frequency range, and detected a 1.5 Jy burst at 50 MHz, whose origin is ambiguous. The source of this emission can be the stellar activity, star-planet interaction, noise fluctuation, or contamination from other sources along the line of sight. A near-contemporaneous, independent NenuFAR analysis by \cite{2026arXiv260708910T} reported a tentative $\sim$10$\sigma$ bursty emission from the same system, although the authors caution that the signal may instead arise from excess correlated noise. HD 189733 Ab has been excluded from our analysis because its host star has an age below 0.7 Gyr, for which the adopted Parker-wind model is not applicable.

$\tau$ Boo Ab, which has the highest predicted flux density in our sample (562.6~mJy at 30.4~MHz, below the 50~MHz lower-frequency limit of SKA-Low; see row 596 of table~\ref{tab1}), yielded a tentative $\sim$3-8$\sigma$ circularly polarized signal in LOFAR beam-formed observations at 14-30~MHz \citep{2021A&A...645A..59T}. The reported flux density of 400-890~mJy is remarkably close to our independent prediction, while the observed frequency range substantially overlaps our predicted characteristic frequency. A dedicated LOFAR follow-up campaign, however, found no confirming signal \citep{2024A&A...688A..66T}, while a deeper imaging observation also resulted in a non-detection \citep{2025A&A...693A.162C}. The apparent agreement between the reported and predicted emission frequencies suggests that $\tau$ Boo Ab may warrant further deep, phase-resolved observations below the SKA-Low frequency range.

Recently, rapid, recurring, highly circularly polarised bursts, along with persistent emission, were reported from $\beta$ Pic b over the frequency range of 0.85-3.5 GHz by \cite{2026arXiv260916720O}. The emission was attributed to the electron cyclotron maser radiation, from which the authors inferred a planetary magnetic field strength of at least 1.25 kG. Despite the significance of this tentative detection, $\beta$ Pic b has not been included in our study because its host star is of spectral type A6V, whereas our sample is restricted to planets orbiting F- to M-type main-sequence stars. Furthermore, the system is very young, with an estimated age of approximately 0.04 Gyr, well below the minimum age adopted in our sample selection.

As an additional comparison with existing low-frequency radio surveys, we assess the detectability of our predicted exoplanet sample using the observing capabilities of the LOFAR Two-metre Sky Survey \citep[LoTSS;][]{2017A&A...598A.104S} and the GaLactic and Extragalactic All-sky MWA survey \citep[GLEAM;][]{2015PASA...32...25W}. We consider systems with characteristic emission frequencies within the LoTSS observing band (120-168~MHz). Adopting the median continuum sensitivity of 92~$\mu$Jy/B from the third LoTSS data release \citep[LoTSS-DR3;][]{2026A&A...707A.198S} and a $5\sigma$ detection threshold, we identify one exoplanet in our sample, HD 89839 b, whose predicted flux density of 0.915~mJy at 126.8~MHz exceeds this threshold. However, HD 89839 b is not observable with LoTSS because its declination (-53.7$\degree$) lies well outside the survey's sky coverage, which extends over 88\% of the northern hemisphere in DR3.

We similarly compare our predicted sample against the observing capabilities of GLEAM. From the full sample of 1027 predicted planets, we select those with characteristic emission frequencies within the GLEAM observing band (72-231 MHz) and declinations below $+30\degree$, corresponding to the survey's sky coverage. Using a representative continuum sensitivity of 10 mJy/B and a $5\sigma$ detection threshold \citep{2017MNRAS.464.1146H}, we find that 33 planets satisfy the band and declination criteria, but none have predicted flux densities exceeding the survey sensitivity; the highest predicted flux among these, GJ 876 b at 20.219 mJy, remains roughly a factor of 2.5 below the GLEAM threshold. Thus, none of the systems in our sample is both sufficiently bright and located within the frequency and sky coverage of either of these two wide-field, low-frequency surveys to have been detected. This further highlights the substantially greater sensitivity and broader frequency coverage of SKA-Low and SKA-Mid, which will open a new observational window for testing these predictions.

\section{Summary and conclusion}\label{conclusion}
We use semi-empirical scaling laws to predict auroral radio emission from confirmed exoplanets orbiting F-M main-sequence dwarf stars and assess its detectability with the SKA during the planned AA4 and intermediate AA* deployment phases. Using parameters from the Extrasolar Planets Encyclopedia, we calculate the radio flux and characteristic emission frequency for 1027 exoplanets using the RBL. Feature-importance analysis via PI and SHAP shows that planetary radius, orbital semi-major axis, stellar age, and Earth-star distance dominate flux density predictions.

Accounting for realistic SKA sky coverage and frequency bands, and adopting conservative 5$\sigma$ imaging sensitivities ($\Delta\nu/\nu_c=0.3$, $\Delta\tau=1$~hour), we find that the AA4 configurations of SKA-Low and SKA-Mid could detect 13 and 2 exoplanets, respectively. In the more compact AA* configuration, 7 SKA-Low targets and both SKA-Mid targets remain detectable. WASP-14 Ab (40.064~mJy at 57.5~MHz) and HD 141937 b (0.084~mJy at 733.1~MHz) are the most promising SKA-Low and SKA-Mid targets, respectively. These results confirm that the SKA can detect gas giants, including WASP-14 Ab with SNR$>$245 and HD 141937 b with SNR$>$19, within a practical integration time. We caution that these thresholds assume nominal Stokes-I sensitivity for a fixed 1-hour integration; the strongly beamed, rotationally modulated, and highly circularly polarised nature of CMI emission means that phase-targeted observations and Stokes-V searches may offer more realistic detection strategies (see section~\ref{obs_cons}).

We also account for the possibility of radio quenching, which can suppress CMI emission when the local plasma density is sufficiently high, particularly for low-mass planets on close-in orbits exposed to strong stellar irradiation. Among the 65 exoplanets predicted to emit within the SKA-Low and SKA-Mid frequency bands, we find that eight lie within the parameter space where radio quenching may be important, including the otherwise detectable SKA-Low candidate TOI-4515 b. This demonstrates that radio-quenching effects should be considered when prioritizing targets for future SKA observations.

The RBL, calibrated solely on the solar system's magnetised planets, carries several structural limitations: its reliance on Blackett-type prescription for the planetary magnetic moment; its inability to capture the Earth-Venus dynamo dichotomy; and its uncertain applicability to super-Earths and sub-Neptunes. Nonetheless, terrestrial-type planets in our sample are predicted to lie below SKA sensitivity thresholds regardless of these uncertainties, leaving gas giants with reliable Jupiter-like analogues as the most robust detectable targets.

Comparison with previous VLA, GMRT, TGSS, and NenuFAR observations shows that several reported radio candidates toward exoplanet-hosting systems remain of uncertain or possibly non-planetary origin. Our predictions suggest that some previously targeted systems are unlikely to be detectable owing to their low expected radio fluxes, while others remain promising targets for future observations (see section~\ref{vla}). Furthermore, none of our predicted planets meets the LoTSS or GLEAM detection requirements.

Given the difficulties in achieving unambiguous detections of exoplanetary radio emission, as evidenced by prior campaigns, advanced instruments like the SKA and next-generation VLA will be crucial for securing definitive results. This underscores the urgent need for predictive models to optimize target selection for upcoming observations. In this context, our study presents a systematic approach for identifying the most promising exoplanetary systems for radio detection with the SKA, demonstrating its transformative potential to detect and characterize exoplanetary radio signals. Such capability opens a new observational window into the magnetic environments of exoplanets and their implications for atmospheric retention and habitability. In addition, this framework can form the basis for future large-scale monitoring programs, helping to allocate observing time more efficiently and improve strategies for detecting magnetospheric interactions in distant planetary systems.

\section*{Acknowledgements}
This research was supported by the Munich Institute for Astro-, Particle and BioPhysics (MIAPbP), which is funded by the Deutsche Forschungsgemeinschaft (DFG, German Research Foundation) under Germany's Excellence Strategy- EXC-2094- 390783311. This research has made use of data obtained from or tools provided by the portal exoplanet.eu of The Extrasolar Planets Encyclopaedia. The authors acknowledge the usage of the \texttt{Scikit-learn} library \citep{2011JMLR...12.2825P}. They also acknowledge the use of OpenAI ChatGPT to improve the readability and clarity of portions of this manuscript.

\section*{Data Availability}
The data underlying this article were derived from Extrasolar Planets Encyclopedia (\url{http://exoplanet.eu/}). The machine-readable versions of the tables are available in the online journal.


\newpage
\appendix
\section{Derivation of RBL}\label{ap1}
The star-planet interaction can be described in terms of the dissipation of the stellar-wind power intercepted by the planetary magnetosphere. In the kinetic-energy formulation of the radiometric Bode's law (RBL), the stellar wind power ($P_{\mathrm{sw}}$) delivered to the planetary magnetosphere is determined by the magnetospheric cross-sectional area and the kinetic energy flux of the stellar wind at the incident point \citep{1984Natur.310..755D,1999JGR...10414025F}:
\begin{equation}
P_{\text{sw}}=m_{p}n(a)v_{\text{eff}}^3\pi R_{m}^2,
\label{a1}
\end{equation}
where $m_p$ is the proton mass, $n(a)$ is the stellar-wind number density at the planetary orbital distance, $v_{\text{eff}}$ is the effective stellar wind velocity in the planetary reference frame, and $R_m$ is the radius of the planet's magnetosphere. $R_m$ is determined from a balance between the planet's magnetic field pressure and the stellar wind's ram pressure as
\begin{equation}
\frac{B^2}{2\mu_0}\left(\frac{R_p}{R_m}\right)^6=m_p n(a)v_{\mathrm{eff}}^2,
\label{balance}
\end{equation}
where $B$ is the planetary magnetic field strength, $\mu_0$ is the vacuum permeability, $R_{p}$ is the radius of planet, and the right-hand side represents the stellar wind ram pressure at the planetary orbital distance. The planet's magnetic field is related to its magnetic moment through $B\propto\mu_{B}/R_{p}^{3}$. Substituting this relation into equation~\ref{balance} yields the magnetopause standoff distance:
\begin{equation}
R_{m}\propto \mu_{B}^{1/3}\left(m_{p}n(a)v_{\mathrm{eff}}^{2}\right)^{-1/6}.
\end{equation}
Here, $\mu_B$ is the planetary magnetic moment, which can be related to the planet's rotation rate ($\omega$) and mass ($M_{p}$) using a Blackett-type scaling law \citep[$\mu_B\propto\omega M_p^{5/3}$,][]{1947Natur.159..658B}. Substituting this relation into equation~\ref{a1} gives
\begin{equation}
P_{\text{sw}}\propto\omega^{2/3}M_{p}^{10/9}n(a)^{2/3}v_{\text{eff}}^{7/3}.
\label{sw}
\end{equation}

To estimate the exoplanetary radio emission properties, we adopt the radiation model of \cite{1999JGR...10414025F}, which assumes that the planet's radio power ($P_{\mathrm{rad}}$) scales with the host star's wind input power as $P_{\mathrm{rad}}\propto P_{\mathrm{sw}}^{1.2}$ \citep{1997pre4.conf..101Z}. Combining this relation with the scaling derived for the stellar-wind input power (equation~\ref{sw}), the planetary radio power can be expressed as
\begin{equation}
P_{\text{rad}}\propto\omega^{0.8}M_{p}^{1.33}n(a)^{0.8}v_{\text{eff}}^{2.8}.
\label{rad}
\end{equation}
The radio power can then be converted into the expected radio flux density at Earth by accounting for the emission bandwidth, beaming geometry, and distance to the observer. The observed flux density is obtained using
\begin{equation}
\Phi=\frac{P_{\text{rad}}}{\Delta f\Omega D^{2}},
\label{flux}
\end{equation}
where $P_{\text{rad}}$ is given by equation~\ref{rad}, $D$ is the Earth-star distance, $\Omega$ is the beaming solid angle of the emission, and $\Delta f$ is the emission bandwidth, which is assumed to be half the characteristic emission frequency, consistent with the observed radio emission of the solar system planets. The characteristic emission frequency ($f_{c}$), which corresponds to the local electron cyclotron frequency in regions of the strongest planetary magnetic field, is
\begin{equation}
f_{c}=\frac{eB}{2\pi m_{e}},
\label{pre_freq}
\end{equation}
where $e$ and $m_{e}$ are the electron charge and mass, respectively. Expressing the planetary magnetic field strength in terms of the planet's radius and magnetic moment, and calibrating the proportionality constant using Jupiter's polar field strength, following equation~4 of \cite{2004ApJ...612..511L}, the characteristic emission frequency of an exoplanet can be estimated by
\begin{equation}
f_{c}=23.5\,\text{MHz}\,\Big(\frac{\omega}{\omega_{J}}\Big)\Big(\frac{M_{p}}{M_{J}}\Big)^{5/3}\Big(\frac{R_{p}}{R_{J}}\Big)^{-3}.
\label{freq}
\end{equation}
Eventually, substituting equations~\ref{rad} and \ref{freq} into equation~\ref{flux} results in the radio flux ($\Phi$) expression,
\begin{equation}
\begin{split}
\Phi = 4.6\,\text{mJy}\,\Big(\frac{\omega}{\omega_{J}}\Big)^{-0.2}\Big(\frac{M_{p}}{M_{J}}\Big)^{-0.33}\Big(\frac{R_{p}}{R_{J}}\Big)^{3}\Big(\frac{\Omega}{1.6\,\text{sr}}\Big)^{-1}\\\Big(\frac{D}{10\,\text{pc}}\Big)^{-2}\Big(\frac{n(a)}{1.34\times10^{4}\,\text{cm}^{-3}}\Big)^{0.8}\Big(\frac{v_{\text{eff}}}{100\,\text{km s}^{-1}}\Big)^{2.8},
\end{split}
\end{equation}
where the beaming solid angle is set to 1.6 steradians, matching Jupiter's decametric emission \citep{1984Natur.310..755D,2004JGRA..109.9S15Z}. In this equation, the reference density $1.34\times10^{4}\,\text{cm}^{-3}$ is chosen so that the leading coefficient stays calibrated to the same $\dot M_{\text{ion}}=10^{-11}\,M_\odot\,\text{yr}^{-1}$, $a=1$~AU, $v=100\,\text{km s}^{-1}$ reference point used in the original formulation \citep[e.g. equation~1 of][]{2018A&A...612A..52O}.

\section{Machine learning-based surrogate method}\label{rf}
We present a reduced-parameter, machine learning-based surrogate model for predicting exoplanetary radio flux densities. Among the various machine learning techniques, the random forest algorithm stands out for its robustness and versatility. As an ensemble method, it builds multiple decision trees during training and combines their predictions to improve the model's performance. Random forest is commonly employed for both regression and classification tasks, as it effectively captures complex relationships in the data and can handle data sets with numerous features and missing values \citep{2001MachL..45....5B}.

For the 1027 exoplanets in our data set, the radio flux density is first computed directly using the RBL (see section~\ref{rbl}). These RBL-derived quantities are then adopted as synthetic training labels to train an independent random forest model for predicting $\Phi$. In other words, the target variable is $\Phi$, derived from the RBL model, while features are $R_{p}$, $a$, $t$, and $D$, selected by PI and SHAP (see section~\ref{key}). This strategy allows the random forest to reproduce the behaviour of the RBL while relaxing its strict input requirements. Note that because the parameters span different ranges, they are transformed into logarithmic space for re-scaling. A conventional 80/20 train-test split is used to evaluate model performance. The hyperparameters of the random forest model are tuned to achieve optimal performance. The accuracy of predictions is evaluated using the root mean square error (RMSE), mean absolute error (MAE), and the coefficient of determination ($R^{2}$) between the random forest outputs and the values obtained from the RBL. These performance metrics are defined in equations~\ref{eq10} to~\ref{eq12}, where $\Phi_{RBL}$ and $\Phi_{RF}$ denote the values of flux density obtained from the RBL and the random forest, respectively. $\Phi_{\text{mean}}$ represents the mean of the $\Phi_{RBL}$ values, and $n$ is the total number of samples. Smaller RMSE and MAE values, along with larger $R^{2}$ values, indicate higher model accuracy.

\begin{equation}
RMSE=\sqrt{\sum_{i=1}^{n}\frac{(\Phi_{RBL}-\Phi_{RF})^2}{n}}.
\label{eq10}
\end{equation}

\begin{equation}
MAE=\frac{1}{n}\sum_{i=1}^{n}\mid \Phi_{RBL}-\Phi_{RF}\mid.
\label{eq11}
\end{equation}

\begin{equation}
R^{2}=1-\frac{\sum_{i=1}^{n}(\Phi_{RBL}-\Phi_{RF})^2}{\sum_{i=1}^{n}(\Phi_{RBL}-\Phi_{\text{mean}})^2}.
\label{eq12}
\end{equation}

\begin{figure}
    \centering
    \includegraphics[width=1\columnwidth]{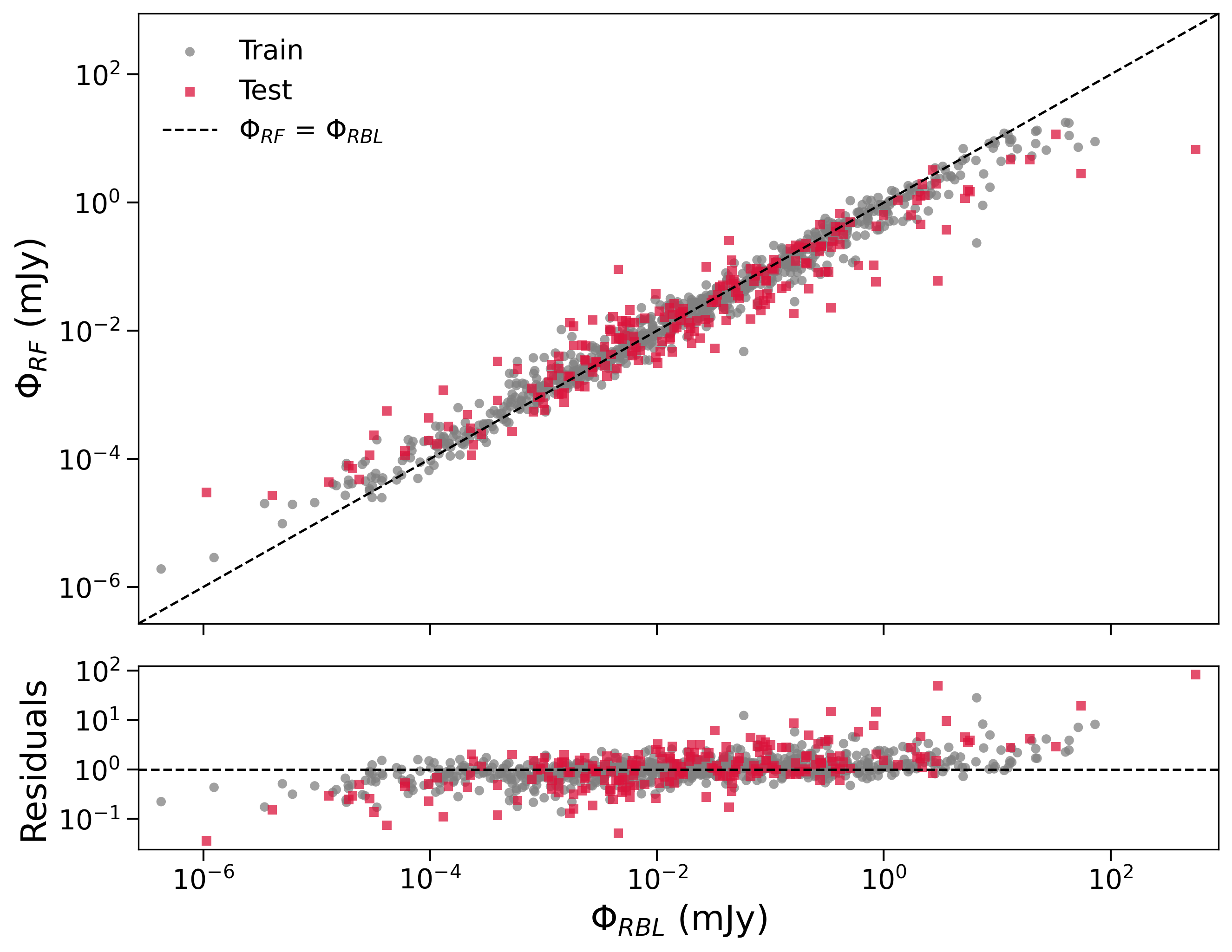}
    \caption{Comparison of the radio flux predicted by the random forest model ($\Phi_{RF}$) with the flux calculated using the radiometric Bode’s law ($\Phi_{RBL}$) in the upper panel, along with the residuals shown in the lower panel. Grey circles represent the training set, while red squares represent the test set.}
    \label{fig_app}
\end{figure}

The random forest model demonstrates strong predictive performance for radio flux, with an RMSE of 0.477, an MAE of 0.340, and a high coefficient of determination ($R^{2}=0.880$) on the test set. To assess potential overfitting in the random forest model, we compare the $R^{2}$ values obtained for the training and test sets. The model achieves an $R^{2}$ of 0.983 on the training set and 0.880 on the test set. The relatively small gap between the two values, together with the high test-set $R^{2}$, indicates good generalisation performance and no strong evidence of substantial overfitting. Figure~\ref{fig_app} compares the radio flux predicted by the random forest model ($\Phi_{RF}$) with the flux calculated using the RBL ($\Phi_{RBL}$). The upper panel illustrates $\Phi_{RF}$ as a function of $\Phi_{RBL}$, with the one-to-one relation ($\Phi_{RF}=\Phi_{RBL}$) marked by the dashed line. The strong alignment of most data points with this line demonstrates that the machine learning model successfully captures and reproduces the expected flux derived from the RBL. The lower panel displays the residuals ($\Phi_{RBL}-\Phi_{RF}$) as a function of $\Phi_{RBL}$, which are centred around zero (horizontal dashed line) over the bulk of the distribution. This indicates that the random forest regression is largely unbiased across the majority of the data. Nevertheless, some deviations are apparent, with the scatter increasing toward both the low- and high-flux regimes, reflecting the limitations of the training set in sparsely populated regions of parameter space.

The residual analysis of the test set across different flux density regimes reveals a clear dependence of the prediction accuracy on flux density. In the intermediate regime (0.0001-1 mJy), the model performs best, with a mean residual close to zero, 0.026, and a standard deviation of 0.38, indicating little evidence of systematic bias for the majority of the test sample. At low flux densities ($<$0.0001 mJy), the mean residual is negative (-0.646), with a standard deviation of 0.35, indicating a tendency to overpredict the faintest sources. Conversely, at high flux densities ($>$1 mJy), the mean residual is positive (0.591) with a standard deviation of 0.54, indicating that the model tends to underpredict the brightest sources. Overall, the residual distribution suggests that the model reproduces intermediate flux densities most reliably, while systematic deviations become more pronounced toward the lower and upper ends of the flux range.

Importantly, reducing the number of input parameters from eight to four does not significantly degrade the random forest model's predictive performance. The test-set RMSE remains essentially unchanged, increasing only from 0.470 to 0.477, while the test-set MAE is nearly identical ($\sim0.340$). Similarly, the test-set $R^{2}$ changes only slightly, from 0.883 for the full model to 0.880 for the reduced model. These results demonstrate that the reduced-parameter model retains almost the same predictive accuracy while requiring only half the number of input parameters. Overall, the radio flux density can be predicted robustly using a substantially simplified parameter set, with the reduced model reproducing the trends captured by the underlying RBL formulation. In this framework, the random forest model is not intended to serve as an independent predictor trained on observational data. Instead, it acts as a surrogate emulator of the RBL, trained on synthetic values generated by the RBL itself. This approach provides an efficient approximation to the mapping between the planetary and stellar parameters and the resulting radio flux densities, while requiring a reduced set of input parameters. In particular, it offers a potential means of estimating radio flux densities for systems with incomplete parameter sets, including those lacking measured planetary masses.

\end{document}